\documentclass[acmsmall,manuscript]{acmart}
\AtBeginDocument{%
  \providecommand\BibTeX{{%
    \normalfont B\kern-0.5em{\scshape i\kern-0.25em b}\kern-0.8em\TeX}}}

\setcopyright{acmcopyright}
\copyrightyear{2018}
\acmYear{2018}
\acmDOI{XXXXXXX.XXXXXXX}
\usepackage{titlesec} 
\usepackage[titletoc]{appendix}
\usepackage{array}
\newcolumntype{C}[1]{>{\centering\arraybackslash}p{#1}}
\usepackage{tikz}
\usepackage{makecell}
\usepackage{graphicx}
\usepackage{adjustbox} 
\usepackage{booktabs}
\usepackage{tabularx}
\usepackage{pdflscape}
\usepackage{enumitem}
\usepackage{longtable}
\usepackage{multirow}
\usepackage{enumitem}
\usepackage{xcolor,colortbl}
\definecolor{lavender}{rgb}{0.9, 0.9, 0.98}
\definecolor{NavyBlue}{RGB}{0, 0, 128}
\definecolor{Cerulean}{RGB}{42, 82, 190}
\definecolor{BurntOrange}{RGB}{204, 85, 0}
\definecolor{OliveGreen}{RGB}{107, 142, 35}
\definecolor{Plum}{RGB}{142, 69, 133}
\definecolor{Goldenrod}{RGB}{218, 165, 32}
\usepackage[edges]{forest}

\usepackage[figuresright]{rotating}
\newcolumntype{g}[1]{>{\centering\arraybackslash\columncolor{lavender}}p{#1}}
\newcolumntype{n}{>{\columncolor{Gray}}c}
\usetikzlibrary{shapes, arrows, positioning, backgrounds}
\titleformat{\section}
{\normalfont\normalsize\bfseries}{\thesection}{1em}{\MakeUppercase}
\acmJournal{JACM}
\acmVolume{37}
\acmNumber{4}
\acmMonth{8}

\usepackage{multicol}
\begin{document}

\title{SoK: Identifying Limitations and Bridging Gaps of Cybersecurity Capability Maturity Models (CCMMs)}

\author{Lasini Liyanage}
\affiliation{
  \institution{University of Auckland}
  \city{Auckland}
  \country{New Zealand}}
\email{lliy930@aucklanduni.ac.nz}

\author{Nalin Asanka Gamagedara Arachchilage}
\affiliation{%
  \institution{University of Auckland}
  \city{Auckland}
  \country{New Zealand}}
\affiliation{%
  \institution{RMIT University}
  \city{Melbourne VIC}
  \country{Australia}}
\email{nalin.arachchilage@rmit.edu.au}

\author{Giovanni Russello}
\affiliation{%
  \institution{University of Auckland}
  \city{Auckland}
  \country{New Zealand}}
\email{g.russello@auckland.ac.nz}

\renewcommand{\shortauthors}{L. Liyanage et al.}

\begin{abstract}

In the rapidly evolving digital landscape, where organisations are increasingly vulnerable to cybersecurity threats, Cybersecurity Capability Maturity Models (CCMMs) emerge as pivotal tools in enhancing organisational cybersecurity posture. CCMMs provide a structured framework to guide organisations in assessing their current cybersecurity capabilities, identifying critical gaps, and prioritising improvements. However, the full potential of CCMMs is often not realised due to inherent limitations within the models and challenges encountered during their implementation and adoption processes. These limitations and challenges can significantly hamper the efficacy of CCMMs in improving cybersecurity. As a result, organisations remain vulnerable to cyber threats as they may fail to identify and address critical security gaps, implement necessary improvements or allocate resources effectively. To address these limitations and challenges, conducting a thorough investigation into existing models is essential. Therefore, we conducted a Systematic Literature Review (SLR) analysing 43 publications to identify existing CCMMs, their limitations, and the challenges organisations face when implementing and adopting them. By understanding these barriers, we aim to explore avenues for enhancing the efficacy of CCMMs, ensuring they more effectively meet the cybersecurity needs of organisational entities. 

\end{abstract}




\maketitle
\section{Introduction}

Cybersecurity has emerged as one of the most pressing concerns for organisations and governments due to the increasing frequency and sophistication of cyberattacks \cite{Alqudhaibi2023}. According to recent reports, over half (53\%) of organisations experienced at least one cyberattack last year, highlighting a widespread vulnerability to threats such as phishing, ransomware, and data breaches \cite{Hiscox2023}. The impact of cyberattacks extends far beyond disruptions of computer systems. These attacks have the potential to interrupt a company's operations for extended periods completely \cite{Aljundi2023}. They also allow hackers to gain unauthorised access to sensitive information, jeopardising customer and proprietary company data \cite{Savacs2022}. For instance, the recent cyberattack on Change Healthcare in the US caused outages in systems used for medical billing and insurance claims, shutting down operations at hospitals and pharmacies for more than a week and gaining access to sensitive patient data \cite{Brooks2024}. Furthermore, the financial repercussions of these breaches can be severe. They include immediate costs such as ransom payments, legal fees, and expenses related to incident response and system recovery \cite{Bellasio2018}. Additionally, there are long-term costs like lost revenue, increased insurance premiums, and reputational damage that can destory customer trust and lead to loss of business \cite{Bellasio2018}. In 2023 alone, the global cost of cyber-attacks reached an estimated \$8 trillion and is projected to rise to \$10.5 trillion by 2025 \cite{Ventures2023}. These figures emphasise the critical need for robust cybersecurity measures to safeguard organisational assets and maintain stakeholder trust.

Despite the evident and ongoing threat of cyber-attacks, many organisations struggle to maintain robust cybersecurity measures. Several key challenges contribute to this struggle: insufficient technological solutions \cite{Agrafiotis2018}, inadequate policies \cite{Mishra2022}, weak procedural safeguards \cite{Maleh2021}, and lacking educational interventions \cite{Shillair2022}. Additionally, the diverse operational models of organisations contribute to the difficulty in adopting uniform cybersecurity strategies \cite{Safitra2023}. 

Addressing these challenges is essential for improving organisational cybersecurity. Cybersecurity Capability Maturity Models (CCMMs) are crucial in this context as they provide a structured framework for assessing and enhancing cybersecurity practices \cite{Buzdugan2023}. CCMMs help organisations systematically identify gaps in their cybersecurity capabilities, prioritize improvements, and implement security best practices \cite{Le2016}. However, the implementation and adoption of CCMMs faces significant obstacles. Existing CCMMs often fail to account for the unique contexts and operational dynamics of different organisations, making them difficult to tailor and implement effectively \cite{Antoniazzi2023}. The complexity of the models, lack of clear guidelines for implementation, and insufficient alignment with organisational priorities further exacerbate the situation \cite{Barnes2022}. Furthermore, many CCMMs overlook critical factors such as organisational culture and resource constraints, leading to resistance and suboptimal adoption \cite{Brown2018}. As a result, these models are often perceived as theoretical and detached from practical realities, resulting in limited practical utility and effectiveness \cite{Buzdugan2022}. The failure to properly implement and adopt CCMMs not only undermines the potential benefits of these models but also keeps organisations stuck with weak cybersecurity measures, leaving organisations vulnerable to evolving cyber threats. Addressing these critical issues is important to enhance the efficacy of CCMMs, ensuring they are adaptable, practical, and aligned with the specific needs and constraints of organisations. By doing so, we can better safeguard organisational assets, maintain stakeholder trust, and improve overall cybersecurity resilience.

Given these significant barriers, it is essential to examine the existing CCMMs and their limitations systematically. 
Conducting a Systematic Literature Review (SLR) allows for a comprehensive analysis of the current body of knowledge. Unlike single empirical studies or expert opinions an SLR allows identifying patterns, gaps, and potential areas for improvement \cite{Awa2017} across a wide range of studies. By critically assessing and integrating findings from various studies, the SLR approach enables us to pinpoint the shortcomings of current models and propose enhancements that are grounded in empirical evidence and tailored to the diverse needs of organisations \cite{Jacobs2017}. Understanding these factors - such as common limitations in current models, recurring implementation challenges, and  influences on CCMM adoption — is vital for making organisations more secure. This knowledge helps to develop more effective and context-specific CCMMs that address the unique challenges and vulnerabilities faced by different organisations. 

To assist organisations and governments in effectively implementing CCMMs and addressing cybersecurity gaps, these models should adopt a holistic perspective that involves organisational, operational, technological and human elements \cite{Marican2023}. This holistic perspective ensures that all aspects of cybersecurity are covered, from aligning practices with organisational goals and integrating them into daily operations, to implementing the right technological tools and addressing human factors through training and awareness. Achieving this involves identifying the existing CCMMs, their limitations, and the challenges faced during their implementation and adoption. By conducting a Systematic Literature Review (SLR) and synthesising findings from various studies, we aim to develop practical recommendations that can be applied across various organisational contexts, thereby enhancing cybersecurity preparedness.

With this objective in mind, we conduct this Systematic Literature Review (SLR) to answer the following key research questions:

\begin{itemize}
    \item[\textit{RQ1.}] \textit{What are the CCMMs introduced to assess cybersecurity capability maturity in organisational entities?}
    \vspace{1em}
    \item[\textit{RQ2.}] \textit{What are the limitations of existing CCMMs encountered by organisations in various contexts when implementing and adopting them?}
    \vspace{1em}
    \item[\textit{RQ3.}] \textit{What are the challenges influencing the implementation and adoption of CCMMs among organisations, considering the diverse industry sectors and organisational settings they operate in?}
\end{itemize}
\vspace{1em}

The rest of the article is organised as follows. Section \ref{Section2-Related Work} briefly describes the related work to this SLR and highlights the research gap and the contribution to knowledge. Next, in Section \ref{Section3-Methodology}, we outline the methodology of the SLR, followed by the findings of the SLR in Section \ref{Section4-Results}. Section \ref{Section5-Discussion} presents the implications of the results. Finally, we discuss some of the limitations
of this SLR in Section \ref{Section6-Limitations} followed by conclusion and future work in Section \ref{Section7-Conclusion}.

\section{Related Work} 
\label{Section2-Related Work}

The evolution of CCMMs has its roots in the broader concept of Information Security Maturity Models (ISMMs), which were initially designed to assess and improve an organisation's information security posture \cite{Konstantinou2020}. These models, inspired by the Capability Maturity Model Integration (CMMI) developed in the early 1990s, provided a framework for organisations to evaluate their processes and practices against established criteria, aiming to enhance their security measures progressively \cite{Yulianto2016, Almuhammadi2017}. The adaptation of ISMMs to cybersecurity came as a natural progression, reflecting the growing recognition of cyber threats and the need to address these challenges in a structured and systematic manner \cite{Abdullahi2020}.

However, the direct application of ISMMs to cybersecurity revealed several limitations \cite{Rabii2020, Yigit2020, Yigit2023, Barnes2022, Spruit2014}. Traditional ISMMs often focused on general information security principles without adequately addressing the specific, dynamic nature of cyber threats \cite{Rabii2020}. They tended to emphasise compliance and process maturity over the agility and adaptability required to counteract evolving cyber threats effectively \cite{Yigit2020}. This gap highlighted the need for more specialised models that could account for the unique challenges of cybersecurity, including the rapid pace of technological change, the sophistication of cyber adversaries, and the increasing complexity of digital ecosystems \cite{Yigit2023, Barnes2022}. As a result, researchers and practitioners recognised the insufficiency of traditional ISMMs for comprehensive cybersecurity management, emphasising the importance of developing models specifically tailored to this field \cite{Spruit2014}.

Cybersecurity maturity models emerged in response to the recognised limitations of ISMMs for comprehensive cybersecurity management \cite{Brown2018}. Targeted efforts have been made to develop cybersecurity maturity models focusing on specific facets of the cybersecurity domain. For instance, models concentrating on cyber counterintelligence have been designed to help organisations assess and enhance their capabilities in identifying, assessing, and countering cyber threats from a strategic intelligence perspective \cite{Jaquire2017, Victor2017, Prasanna2021, Sebastiaan2017}. Similarly, cyber resilience maturity models have emerged, focusing on enhancing an organisation's ability to anticipate, withstand, recover from, and adapt to adverse cyber events or conditions \cite{Baikloy2020, Ebrahimnezhad2023, Shaked2020, Keleba2022, Kulugh2022, Schlette2021, Sharkov2020}. Additionally, models addressing cyber hygiene have been developed to focus on the foundational practices and controls that ensure the basic health and security of an organisation's digital infrastructure \cite{Skarga2021}. These models, while vital in addressing particular areas of cybersecurity, they do not provide organisations a roadmap to achieve overall cybersecurity maturity. 

Amidst more niche models like those focusing on cyber counterintelligence, cyber resilience, or cyber hygiene, Cybersecurity Capability Maturity Models (CCMMs) were introduced, offering an evaluation of an organisation's overall cybersecurity capability maturity \cite{Malatji2019}. CCMMs do not merely address specific threats or resilience strategies but evaluate the maturity of an organisation's cybersecurity capabilities as a whole \cite{Imran2022}. This assessment is essential in today’s digital age, where cyber threats are multifaceted and rapidly evolving. Without a comprehensive understanding of their cybersecurity posture, organisations risk overlooking critical vulnerabilities that adversaries could exploit, leading to breaches that compromise sensitive data, disrupt operations, and destroy stakeholder trust \cite{Aliyu2020}.

To achieve this comprehensive evaluation, CCMMs encompass three core components: capability domains, practices and maturity levels \cite{Almuhammadi2017}. The capability domains define key areas of cybersecurity focus (e.g. risk management, identity and access management, asset management, training and awareness) \cite{Karabacak2016}. They help organisations ensure that all essential areas are covered and reduce the risk of oversight due to inadequate defences. Within each capability domain, CCMMs provide a set of actionable practices. These are detailed, practical guidelines that help organisations implement effective cybersecurity measures to enhance cybersecurity across technological, operational, and educational domains \cite{Barclay2014}. CCMMs also outline progressive maturity levels that guide organisations through incremental stages of cybersecurity development \cite{Moller2023}. This structured progression helps organisations systematically enhance their cybersecurity maturity from basic to advanced levels \cite{Karabacak2016, Drivas2020}. Therefore, together, these components create a roadmap that organisations can use to strengthen their cybersecurity, prioritise investments, and benchmark their progress against industry standards \cite{Selamat2022, Brown2018, Adler2013}.

There are Generic CCMMs \cite{C2M2, Walter2014, Ghaffari2018, Dube2020} and Specific CCMMs \cite{ONGC2M2, ESC2M2, Kour2020, Gourisetti2020, White2011, Von2015, Akinsanya2019, Gourisetti2019, Drivas2020, Daraghmeh2021, Bhattacharya2022, Kreppein2021, Adler2013, Aliyu2020, Karabacak2016, Selamat2022, Le2017, Rojas2022, Barclay2014, Brown2018} developed by both industry and academia. Generic CCMMs offer a wide-ranging framework applicable across various sectors, aiming to provide a foundational approach to assessing and enhancing cybersecurity posture \cite{Moller2023}. On the other hand, specific models are tailored to address the requirements of particular industries or sectors, offering more targeted insights \cite{Moller2023}. However, significant shortcomings can be seen in their design and application \cite{Rabii2020}. Many of these models lack specificity and are overly broad, making it difficult to contextualise them to unique organisational environments \cite{Anne2018}. For example, a technology startup might find it difficult to implement generic CCMM guidelines due to limited resources and need for agile flexible security measures tailored for their fast-paced environment \cite{Marican2023}. Additionally, specific models, while more targeted, often fail to encompass the full range of cybersecurity concerns within their narrow focus. For instance, a model designed for the financial sector might overlook critical elements like physical security controls that are crucial in sectors such as healthcare, manufacturing, and critical infrastructure \cite{Anne2018}. Additionally, a model designed for the financial sector might overlook critical elements like physical security controls that are crucial in other sectors such as healthcare, manufacturing, and critical infrastructure, demonstrating how specific models, while more targeted, often fail to encompass the full range of cybersecurity concerns within their narrow focus \cite{Rojas2022, Karabacak2016}. 

Furthermore, the effectiveness of implementing and adopting CCMMs across diverse organisational contexts remains under-explored, leading to challenges in their effective application \cite{Brown2018}. While studies have been conducted to develop both generic models and models for various sectors (i.e. healthcare, finance, transport), the specific challenges faced by small and medium-sized enterprises (SMEs) and large-scale enterprises in implementing and adopting these models are largely unexplored. SMEs often lack the resources and expertise needed to effectively implement comprehensive CCMMs, while large-scale enterprises may struggle with integrating these models into their complex, multi-layered structures. This gap in understanding results in organisations missing opportunities to adapt CCMMs to their unique operational contexts and effectively utilise emerging technologies \cite{Anne2018}. Therefore, the right balance between providing a comprehensive assessment and ensuring its accessibility and customisation to meet organisational needs should be achieved. To accomplish that, the limitations of the existing CCMMs and challenges associated with implementing and adopting CCMMs should be identified.

Nevertheless, to date, no publication has primarily analyzed the literature on CCMMs, their limitations, and challenges for implementation and adoption through a Systematic Literature Review (SLR). This gap may be attributed to several factors. First, the complexity and diversity of organisational contexts make it challenging to develop a one-size-fits-all approach \cite{Ghaffari2018}, resulting in fragmented research. Second, the evolving nature of cybersecurity threats necessitates continuous updates to CCMMs \cite{Buzdugan2023}, which can be difficult to capture comprehensively in a single study. Finally, the wide variety of CCMMs developed by both industry and academia adds to the complexity, as each model may focus on different aspects of cybersecurity.

For example, Rabii et al. \cite{Rabii2020} compared the structures of ISMMs alongside CCMMs to understand their shared features and implementation pathways. Marican et al. \cite{Marican2023} assessed CCMMs as a framework for cybersecurity maturity assessment with other ISMMs and security frameworks (i.e., ISO 27001, ISO 21827, Cloud Security Alliance, COBIT) to determine their suitability for implementation within technology startups. While these studies provide valuable insights, they are limited in scope and do not offer a comprehensive synthesis and critique of a broader range of CCMMs across different organisational contexts and industries.

The lack of a comprehensive SLR leaves a significant gap in understanding the overarching trends, recurring challenges, and potential improvements needed for CCMMs. An SLR is essential in this case because it systematically reviews and integrates findings from a diverse body of literature, providing a holistic overview of available CCMMs, their limitations, and the challenges influencing their implementation and adoption across various sectors. This approach ensures that the review captures the full spectrum of existing knowledge and identifies gaps that individual studies may not reveal.

Therefore, this Systematic Literature Review (SLR) aims to fill this gap by providing a holistic overview of the available CCMMs for organisations to adopt, identifying and synthesising the limitations of these models as reported in the literature, and exploring the challenges that influence their implementation and adoption across various sectors. By doing so, this review will offer a comprehensive analysis that incorporates a broader range of models, including those not previously covered in-depth, and assess their effectiveness, limitations and challenges in a wider array of organisational contexts. Policymakers, industry experts, and cybersecurity practitioners can draw on these insights to develop tailored strategies, best practices, and recommendations for optimising these models and maximising the benefits of these frameworks. 

\section{Methodology}
\label{Section3-Methodology}

This study is undertaken as an SLR based on the guidelines proposed by Kitchenham and Charters \cite{Kitchenham02, Santos2019}, complemented by a thematic analysis \cite{Braun2012} to facilitate the qualitative analysis and synthesis of the identified research articles. The primary objective of this review is to identify and analyse the prior studies in CCMMs employed within organisations to uncover their inherent limitations and challenges for implementation and adoption, ultimately aiming to enhance their effectiveness and applicability across diverse organisational contexts.

The review methodology consists of two distinct phases \cite{Kitchenham04}: (1) Planning the review and (2) Conducting the review, which is discussed below. 

\subsection{Planning the review}

Planning the review phase involves the preparatory steps essential for conducting a systematic and effective literature review. It consists of the following activities \cite{Kitchenham04}.

\begin{enumerate}
    \item Defining the scope of the research
    \item Formulating the research questions
    \item Developing the search string
\end{enumerate}

\subsubsection{Defining the scope of the research}\hfill

\hspace{-1.2em}
This study uses the PICOC (Population, Intervention, Comparison, Outcome, Context) framework to define the scope of the research. The PICOC framework provides a structured and holistic approach to delineating the essential components of the research study \cite{Kitchenham2013}. A summary of the framework is demonstrated in Table \ref{tab:PICOCModel}.

The population of interest in this review are the organisations implementing and adopting CCMMs. Hence, the interventions under scrutiny are the existing CCMMs. Once the existing maturity models are identified, the study compares and evaluates the models to identify their limitations and challenges for implementation and adoption. This study is conducted within the context of multifaceted industry sectors and organisational settings which confront diverse cybersecurity challenges and endeavour to improve their cybersecurity practices. By addressing these elements - population, intervention, comparison, outcome, and context - the PICOC framework aids in shaping the research scope by ensuring that the research is well-structured and relevant \cite{Booth2012}.

\begin{table*}[htbp]
\begin{tabular}{p{5em} p{30em}}
\toprule
\textbf{Criterion} & \textbf{Application} \\
\midrule
    Population & \underline{Organisations} that are implementing and adopting CCMMs \\
    \vspace{0.1em}
    Intervention & \vspace{0.1em} \underline{Cybersecurity capability maturity models} \\
    \vspace{0.1em}
    Comparison & \vspace{0.1em} \underline{Comparison} and \underline{evaluation} of the existing CCMMs \\
    \vspace{0.1em}
    Outcome & \vspace{0.1em} Identification of the \underline{limitations} and \underline{gaps} in the existing models and \underline{challenges} for \underline{implementation} and \underline{adoption} \\
    \vspace{0.1em}
    Context & \vspace{0.1em} Diverse \underline{industry sectors} and \underline{organisational settings} facing cybersecurity challenges and seeking to enhance their cyber security practices \\
\bottomrule
\end{tabular}
\caption{Research scope defined based on the PICOC Framework}
\label{tab:PICOCModel}
\vspace{-2.5em}
\end{table*}

\subsubsection{Formulating the research questions}\hfill
\label{RQFormulation}

\hspace{-1.2em}
Initially, the Research Questions (RQs) were identified based on the objectives of the study. These questions were then refined using the distinct elements of the PICOC framework to ensure comprehensive coverage of the relevant aspects. In RQ1, the question maintains its focus on the availability of CCMMs, directly addressing the "Intervention" component by inquiring about models for organisations to adopt. RQ2 inquires about the "limitations" encountered by organisations when implementing and adopting these models "in various contexts" by directly encompassing the "Outcome" and "Context" aspects. RQ3 addresses both "factors and challenges" and "implementation and adoption" directly, aligning it with the "Intervention" and "Outcome" components while accounting for the varying organisational contexts and settings as highlighted in the "Context" element. These questions not only sharpen the focus of the research by providing clear and specific areas of inquiry but also ensure that it explicitly encompasses the PICOC criteria. This enhances the relevance and applicability of the study \cite{Kitchenham2013} by systematically addressing the key components critical to understanding the effectiveness of CCMMs. 
\vspace{1em}

\begin{itemize}
    \item[\textit{RQ1.}] \textit{What are the CCMMs introduced to assess cybersecurity capability maturity in organisational entities?}
    \vspace{1em}
    \item[\textit{RQ2.}] \textit{What are the limitations of existing CCMMs encountered by organisations in various contexts when implementing and adopting them?}
    \vspace{1em}
    \item[\textit{RQ3.}] \textit{What are the challenges influencing the implementation and adoption of CCMMs among organisations, considering the diverse industry sectors and organisational settings they operate in?}
\end{itemize}
\vspace{1em}

\subsubsection{Identifying the Search Terms}\hfill

\hspace{-1.2em}
The next step of the review process is identifying the search terms required to develop the search string that aligns with the research scope and the RQs to retrieve the relevant literature \cite{Kitchenham2013}. As demonstrated in Table \ref{tab:PICOCModelSearchTerms}, the selection of search terms has been tailored to align with the identified PICOC criteria in Table \ref{tab:PICOCModel} and the improved research questions. For the "Population" component, a comprehensive list of synonymous terms such as "organisation," "institution," "firm," and "government" has been included to encompass a broad spectrum of entities that might be implementing CCMMs. Likewise, the "Intervention" component is thoroughly covered with terms like "cybersecurity capability maturity model," "maturity model," and various permutations of "cybersecurity" and "cyber security," ensuring that all relevant models are considered. The "Comparison" terms such as "comparison," "evaluation," and "analysis" directly address the need to assess and contrast these models, as specified in the research questions. The "Outcome" terms, including "gaps," "limitations," and "challenges," mirror the research focus on identifying issues and factors related to model implementation and adoption, which are central to the research questions. Lastly, the "Context" terms extend the search to diverse industry sectors and organisational settings, encompassing fields like "healthcare," "finance," and "defence," thereby aligning with the research's emphasis on examining these models in various real-world contexts. Additionally, synonyms and variations (i.e. inflectional variations, derivational variations, spelling variations) of the derived search terms are incorporated to broaden the search scope as they account for potential linguistic differences in the academic literature \cite{Wolfram2021}. 

\begin{table*}[htbp]
\begin{tabular}{p{5em}p{30em}}
\toprule
\textbf{Criterion} & \textbf{Search Terms} \\
\midrule
    Population & organisation(s), institution(s), firm(s), enterprise(s), institute(s), organisation(s), company(ies), business(es), corporation(s), government(s) \\
    \vspace{0.1em}
    Intervention & \vspace{0.1em}  cybersecurity capability maturity model(s), cyber security capability maturity model(s), cybersecurity maturity model(s), cyber security maturity model(s),  maturity model(s), capability maturity model(s) \\
    \vspace{0.1em}
    Comparison &  \vspace{0.1em} comparison, evaluation, analysis, assessment, compare, evaluate, analyze, analyse, assess \\
    \vspace{0.1em}
    Outcome & \vspace{0.1em}  gaps, limitations, drawbacks, issues, barriers, challenges, factors, adoption, adopt, implementation, implement \\
    \vspace{0.1em}
    Context & \vspace{0.1em} industry(ies), industrial, sector(s), domain(s), public, private, critical infrastructure, cybersecurity, cyber security, cybersecurity posture \\
\bottomrule
\end{tabular}
\caption{Search terms defined based on the PICOC Framework}
\label{tab:PICOCModelSearchTerms}
\vspace{-2.5em}
\end{table*}

\subsubsection{Developing the Search String}\hfill

\begin{figure}[htbp]
    \centering \includegraphics[width=\textwidth]{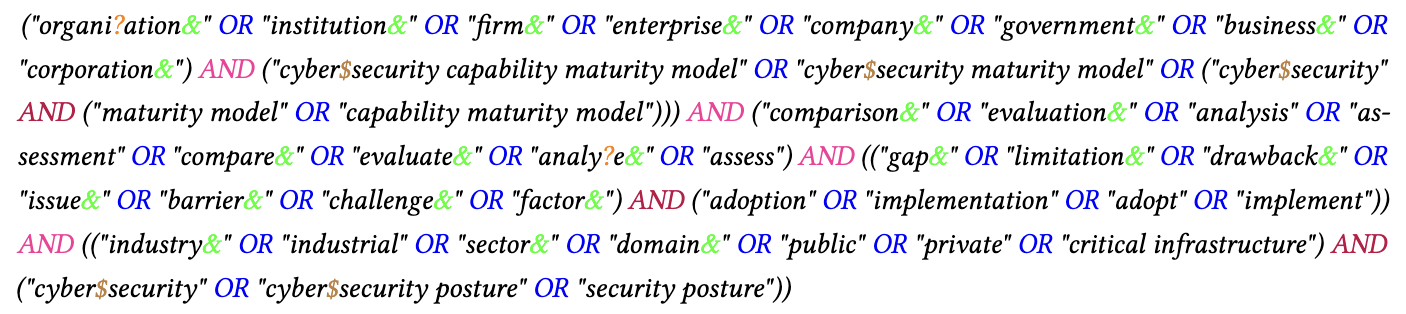}
    \caption{Search String}
    \label{fig:SearchString}
\end{figure}

\hspace{-1.2em}
To create the search string, the search terms identified for each PICOC criterion in the previous step are combined using Boolean operators and truncation symbols \cite{Kitchenham02} as shown in \ref{fig:SearchString}. The terms associated with Population, Intervention, Comparison, Outcome, and Context are connected with 'AND' operators to create a logical link that narrows the search to encompass the essential elements of the research focus. 

The use of 'OR' with terms like "organisation," "institution," "firm," and so forth ensures that the search encompasses a wide range of synonymous population descriptors \cite{Kitchenham02}. The "?" and "\&" truncation symbols with terms like "organi?ation\&" accommodates spelling variations with zero or one character and singular and plural forms of the words \cite{Jones2004}, respectively. In the "Intervention" component, the 'OR' and 'AND' operators are used to combine terms related to CCMMs while utilising "\$" to account for variations in writing with one or no character \cite{Jones2004}. Within the "Comparison" element, the list of related terms emphasising the evaluation and analysis of the models are connected by 'OR' operators. 

Similarly, within the "Outcome" component, the 'OR' operator combines terms related to limitations and challenges, and the terms related to adoption and implementation are connected using the 'AND' operator. Lastly, for the "Context" component, the 'OR' operator effectively combines industry sectors, and the 'AND' operator combines the settings with cybersecurity-related terms, encompassing the diverse contexts specified in the PICOC criteria. The careful use of Boolean operators and truncation symbols, as discussed, ensures a robust search that covers a broad spectrum of relevant literature while maintaining specificity.

Similarly, within the "Outcome" component, the 'OR' operator combines terms related to limitations and challenges, and the terms related to adoption and implementation are connected using the 'AND' operator. Lastly, for the "Context" component, the 'OR' operator effectively combines industry sectors. The 'AND' operator combines the settings with cybersecurity-related terms, encompassing the diverse contexts specified in the PICOC criteria. The careful use of Boolean operators and truncation symbols, as discussed, ensures a robust search that covers a broad spectrum of relevant literature while maintaining specificity.

The planning phase involves defining the research scope using the PICOC framework, formulating research questions, and developing the search string with appropriate search terms. This structured approach ensures a comprehensive and focused review process. The next phase, conducting the review, will methodically explore the existing literature to gather, evaluate, and organise relevant information.

\subsection{Conducting the review}

This phase involves methodical exploration of the existing literature, focusing on gathering, evaluating, and organising relevant information \cite{Kitchenham04}. It encompasses the following steps to ensure a thorough and unbiased review process.

\begin{enumerate}
    \item Search strategy
    \item Study selection
    \item Data extraction
    \item Data analysis
\end{enumerate}

\subsubsection{Search Strategy}\hfill

\hspace{-1.2em}
The search strategy incorporates the established search string in conjunction with the PRISMA (Preferred Reporting Items for Systematic Reviews and Meta-Analyses) \cite{Page2021} flow diagram (Figure \ref{fig:PRISMAFlowChart}). The PRISMA flow diagram was employed to visually represent the search process, illustrating the flow of studies from initial identification through to final inclusion \cite{Kitchenham02}. The diagram depicts the stages of study selection, highlighting the number of articles identified, screened, excluded, and ultimately included in the review \cite{Page2021}. 

In phase 1, prominent online databases and literature search engines were systematically searched using the developed search string to identify the relevant literature. The selected online databases and search engines were, IEEE Xplore\footnote{\url{https://ieeexplore.ieee.org/}}, ACM Digital Library\footnote{\url{https://dl.acm.org/}}, Science Direct\footnote{\url{https://www.sciencedirect.com/}}, Springer Link\footnote{\url{https://link.springer.com/}}, Scopus\footnote{\url{https://www.scopus.com/}}, and Google Scholar\footnote{\url{https://scholar.google.com/}}.

The search string was adjusted iteratively to optimise the literature search's effectiveness in retrieving relevant materials and achieving comprehensive coverage through several pilot searches \cite{Van2021}. For instance, if a pilot search yielded an overwhelming number of irrelevant results, Boolean operators or truncation symbols were modified to make the search more precise. Conversely, if the search retrieved too few relevant articles, the search string was expanded by including additional synonyms or variations of search terms. The adjustments were made with careful consideration of the database's unique syntax, handling of Boolean operators, and truncation symbols, ensuring that the search string remains compatible and effective for each specific platform \cite{ColoradoState2017}.

In addition to the online databases, a tailored search strategy was implemented to identify relevant articles from top-tier peer-reviewed conferences and journals \cite{Kitchenham04} in Phase 2. Since this study straddles the domains of cybersecurity and information systems due to its inherent interdisciplinary nature, renowned conferences and journals in cybersecurity and information systems disciplines were considered. The study aligns with the cybersecurity domain, delving into assessing and enhancing organisations' cybersecurity practices, focusing on their maturity levels. On the other hand, the study also intersects with the information systems domain because it investigates how organisations manage and leverage their IT and information systems to achieve cybersecurity maturity. 

We considered 9 conferences and 9 journals to search publications, which are detailed in Table \ref{tbl:Conferences and Journals}. These conferences and journals were chosen for their reputation for hosting and publishing cutting-edge cybersecurity and information systems research, ensuring the review included the most current and authoritative insights. The tailored search across these venues complemented the database searches and enhanced the likelihood of capturing relevant articles that might have been missed initially \cite{Kitchenham04}.

\begin{table*}[htbp]
\begin{tabular}{  p{9em} p{17em} p{17em} }
\toprule
\textbf{Track} & \textbf{Conference} & \textbf{Journal} \\
\midrule
  \multirow{7}{5cm}{Cybersecurity} & IEEE Symposium on Security and Privacy (IEEE S\&P) & ACM Transactions on Privacy and Security (TOPS) \\ 
   & USENIX Security Symposium & Journal of Cybersecurity (J. Cybersecur.) \\
   & ACM Conference on Computer and Communications Security (CCS) & Computers \& Security \\
   & Symposium on Usable Privacy and Security (SOUPS) &  \\ 
  \hline                                                    \multirow{7}{5cm}{Information Systems} & European Conference on Information Systems (ECIS) & European Journal of Information Systems (EJIS) \\
   & International Conference on Information Systems (ICIS) & Information Systems Journal (Inf. Syst. J.) \\
  & Americas Conference on Information Systems (AMCIS) & Journal of Strategic Information Systems (J. Strateg. Inf. Syst.) \\
  & Pacific Asia Conference on Information Systems (PACIS) & Information Systems Research (Inf. Syst. Res.) \\ 
  & Hawaii International Conference on System Sciences (HICSS) & Journal of the Association for Information Systems (JAIS) \\
   &  & MIS Quarterly (MIS Q.) \\ 
\bottomrule
\end{tabular}
\caption{Selected Conferences and Journals}
\label{tbl:Conferences and Journals}
\vspace{-3em}
\end{table*}

Furthermore, a backwards snowballing approach was employed in Phase 3 to support the automated search processes followed in Phase 1 and 2. This technique involves reviewing the reference lists of identified articles in Phase 1 and 2 to discover additional studies that might be of interest \cite{Van2021}. Combining these diverse search methods, the literature search process was executed, enabling the retrieval of a comprehensive set of materials for the systematic literature review. 

\subsection{Study Selection}
\begin{figure*}[htbp]
    \centering \includegraphics[scale=2, width=\textwidth]{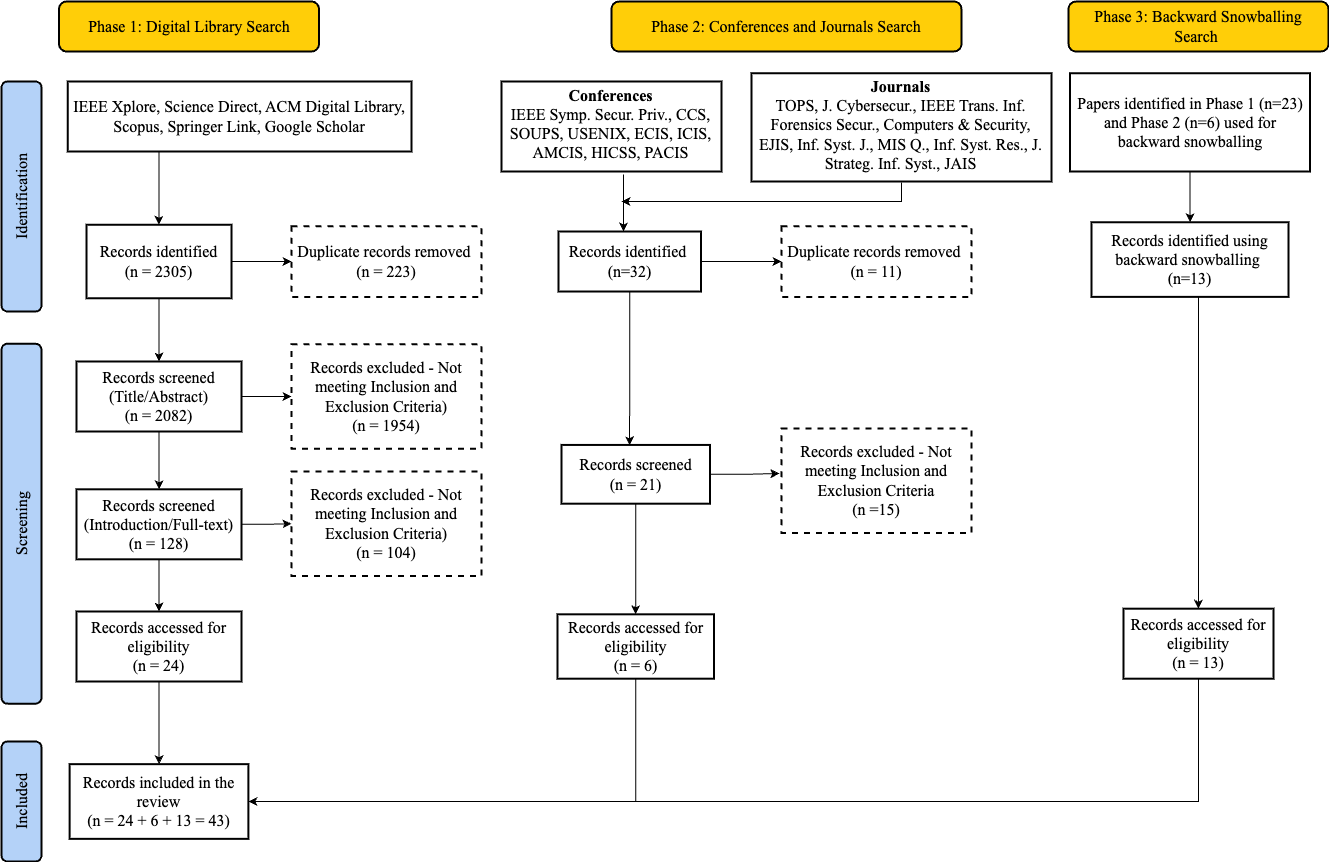}
    \caption{PRISMA Flow Diagram for Study Selection}
    \label{fig:PRISMAFlowChart}
\end{figure*}

The study selection procedure followed in this study is shown in the PRISMA flow diagram in Figure \ref{fig:PRISMAFlowChart}. A total of 2305 papers from January 2011 to April 2024 were identified using the automated search using online databases. 223 duplicates were removed from the initially identified articles, where 2082 individual articles were selected for further screening. Among them, 1954 studies were excluded by going through their titles and abstracts as they did not meet the inclusion and exclusion criteria in Table \ref{tbl:InclExclCri}. We defined the Inclusion Criteria (IC) and Exclusion Criteria (EC) to determine which studies should be included and excluded from the SLR \cite{Fink2019, Kitchenham2013} as per the scope defined in Table \ref{tab:PICOCModel}.

We added the remaining 128 publications that met the inclusion and exclusion criteria to Zotero Reference Management Software\footnote{\url{https://www.zotero.org/}}. They were reviewed further by going through their abstracts, introduction and full-text and 24 publications meeting the aforementioned criteria were selected for the study in Phase 1. 

Upon conducting the search within reputed conferences and journals listed in Table \ref{tbl:Conferences and Journals}, a total of 32 articles were identified in Phase 2. After removing the duplicates and irrelevant publications, 6 publications eligible for the review were selected. 

The backward snowballing technique was applied to the selected articles identified from online databases and reputed conferences and journals, and a further set of 13 related literature was identified. Following this screening process, at the end of the three phases, 43 papers were selected for SLR.

\begin{table*}[htbp]
\begin{tabular}{  p{2em}  p{35em} } 
\toprule
\textbf{ID} & \textbf{Inclusion Criteria} \\
\midrule
  IC1 & Publications that propose, apply, validate, or analyse one or more CCMMs  \\
  IC2 & Peer-reviewed journal articles, conference proceedings, technical reports, workshop papers, and reputed books \\ 
  IC3 & The most recent version of the publication, if it has previously published versions \\
  IC4 & Publications written in English \\ 
  \hline
   & \textbf{Exclusion Criteria} \\
\midrule
  EC1 & Publications that propose or apply a method for measuring an organisation’s level of cybersecurity maturity but do not refer to a cybersecurity capability maturity model\\
  EC2 & Publications that focus on maturity models for cyber counterintelligence, cyber hygiene, cyber risk management, cybersecurity capacity, cybersecurity governance, and cybersecurity threat \\
\bottomrule
\end{tabular}
\caption{Inclusion and Exclusion Criteria}
\label{tbl:InclExclCri}
\vspace{-2em}
\end{table*}

\subsection{Data Extraction}
Data extraction was carried out to collect data items that are suitable for answering research questions. The full texts of the 43 publications identified through study selection were read, and the essential data for our study was collected using a data extraction form. 

A base extraction form was established using the research questions. Afterwards, a pilot extraction was carried out to update the data extraction form and iteratively update the data extraction form to incorporate the relevant data elements that need to be collected \cite{Kitchenham04}. Following several iterations of selecting a limited number of studies and changing the data extraction process, 10 data extraction elements, namely Type of the study, Scope of the study, Objective of the study, Research Gaps, CCMMs covered, Evaluation performed, Summary of main results, Limitations \& Gaps of the models, Factors \& challenges for adoption \& implementation, Conclusion and Future work were finalised. This form also included general metadata such as title, authors, publication, and year of publication. This data form was implemented in MS Excel, allowing further data synthesis to identify patterns.

\subsection{Data Analysis}
\label{DataAnalysisSection}

We aim to answer the research questions through a qualitative exploration by identifying and interpreting recurring themes and patterns within the gathered data. Therefore, we conducted a Reflexive Thematic Analysis \cite{Braun2012} adhering to the following six-phase process. Reflexive Thematic Analysis helps researchers to engage with their own knowledge, experiences and and perspective to delve deeper into the data and extract meaningful patterns in them. This method particularly suits our study due to the diverse expertise of the authors. All authors have industry experience in cybersecurity and information systems and some are well-established researchers in the implementation and adoption of cybersecurity interventions in organisations. This reflexivity offered by this combination helps in leveraging practical insights with theoretical understanding ensuring comprehensive exploration of CCMMs in different organisational contexts.

\subsubsection{Familiarisation with the data.} 
Extracted data was read and summarised using the Zotero Reference Management Software to have an overview of the data.

\subsubsection{Generating codes.} 
We adopted the integrated coding approach to generate the shorthand descriptive labels (codes) for the pieces of information relevant to the research questions (i.e., RQ1, RQ2, and RQ3) \cite{Cruzes2011}. The authors first created a provisional 'start list' of codes based on the research questions and the authors' knowledge of CCMMs (deductive approach). Then, the authors expanded the code list by adding new codes based on the extracted data (inductive approach). 

For example, the 'start list' of codes consisted of codes such as "too complex", "inflexible", and "resource intensive" to reflect the limitations in maturity models concerning RQ2. The list was expanded by assigning codes such as "compliance vs maturity" and "inadequate maturity metrics" to the texts "cyber security maturity models should focus on more than standards compliance" \cite{Drivas2020} and "most cyber security maturity models draw on security practices measured by qualitative metrics/processes; quantitative metrics should be essential for any security assessment." \cite{Le2017} identified in the data respectively. 

\subsubsection{Generating themes.}
In this step, the authors compiled all the codes extracted from the data into logical groups to identify themes that help to answer the research questions. For instance, the authors grouped the codes such as "too complex" and "inflexible" related to the limitations of the CCMMs into the "Technical" theme (described in Section \ref{Limitations}), which highlights technical limitations concerning the adaptability and flexibility of the maturity models. 

\subsubsection{Reviewing potential themes.}
The authors then cross-referenced these themes with the data segments extracted from the literature to ensure they collectively formed a coherent narrative about existing CCMMs, their limitations and challenges. At times, the authors refined the initial themes to make them more specific and representative of the meanings within the data. For example, the authors initially grouped various challenges under a broad theme like "implementation challenges," but upon closer examination, we split this theme into more specific categories, such as "organisational challenges" and "technical challenges".

\subsubsection{Defining and naming themes.}
Finally, we defined two main themes as generic (non-sector specific) CCMMs and sector-specific CCMMs, to answer the RQ1, where the domains of the models were further analysed as more detailed themes. In order to answer the RQ2, the limitations of the models identified through the data were categorised into three themes, namely Social, Technical and Environmental.\\

The co-authors cross-verified the coding process to ensure the alignment of codes and themes with the underlying data, as well as their relevance to the research questions RQ1, RQ2 and RQ3. This was achieved through comprehensive discussions among the three authors in multiple sessions aimed at resolving any discrepancies in the coding and thematic development. This collaborative approach significantly reduced potential coding inconsistencies. After creating themes and categories, the main author categorised the research publications using a spreadsheet based on the detailed descriptions of these themes and categories. In subsequent discussions, the coding team articulated and documented their consensus and dissent on the categorisation choices. The authors placed greater emphasis on integrating the diverse viewpoints of the co-authors in the thematic development and the assignment of publications, in line with the guidance provided by Braun and Clark \cite{Braun2012}.

\section{Results}
\label{Section4-Results}

The absence of a comprehensive CCMMs that can be adopted across various organisational contexts presents a significant challenge for organisations and governments in accurately evaluating their cybersecurity capabilities \cite{Buzdugan2023}. This gap often results in organisations struggling to fully understand and assess their cybersecurity capabilities. One of the primary reasons for this struggle is the inherent complexity and lack of adaptability of existing CCMMs, which may not align well with the unique needs and contexts of different organisations \cite{CybermaturityISACA}. The lack of a such comprehensive and flexible model makes it difficult for organisations to implement and adopt CCMMs effectively. Consequently, they may overlook critical security controls, leaving themselves exposed to various cyber threats with potentially severe consequences \cite{CSStrategy, Walter2014}. To mitigate these risks, it is essential to address the inherent limitations in the existing models and challenges associated with the implementation and adoption of CCMMs, and to develop strategies for creating a more effective and adaptable model that organisations can reliably use to enhance their cybersecurity maturity.

\begin{figure}[h!]
    \centering \includegraphics[height=15cm, width=17cm]{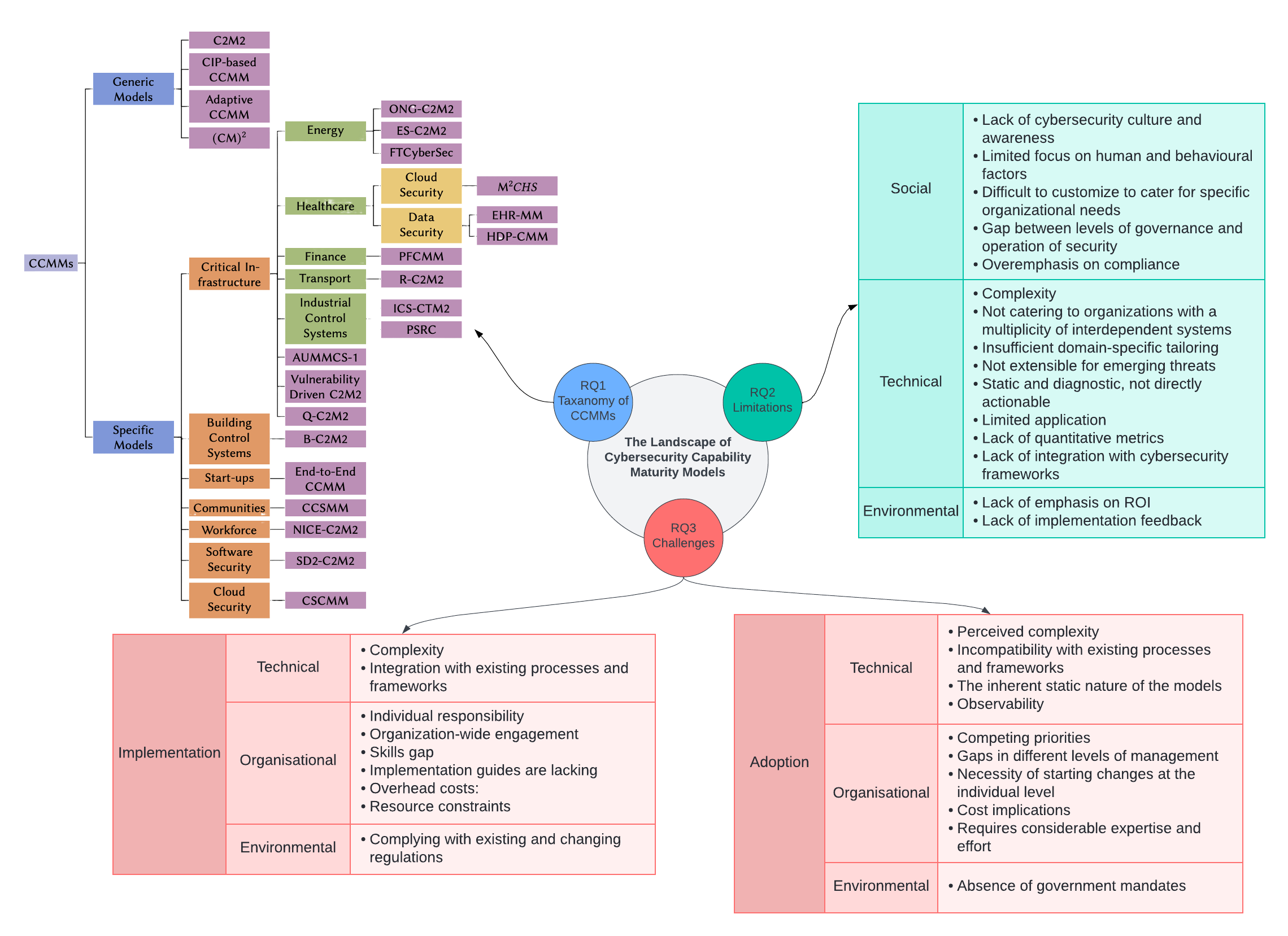}
    \caption{The overview of the themes identified from the reflexive thematic analysis of the SLR and the main findings under each theme. The themes are categorised under respective research questions. RQ\# = Research Question}
    \label{fig:Results}
\end{figure}

With that motivation, we conducted this SLR identifying two primary categories of CCMMs, namely Generic Models \cite{C2M2, Walter2014, Ghaffari2018, Dube2020} and Specific Models \cite{ONGC2M2, ESC2M2, Kour2020, Gourisetti2020, White2011, Von2015, Akinsanya2019, Gourisetti2019, Drivas2020, Daraghmeh2021, Bhattacharya2022, Kreppein2021, Adler2013, Aliyu2020, Karabacak2016, Selamat2022, Le2017, Rojas2022, Barclay2014, Brown2018}, their inherent limitations and challenges for the implementation and adoption as shown in Figure \ref{fig:Results}. Identifying those limitations and challenges is imperative to refine existing models and assist organisations in selecting and adapting a maturity model that aligns with their unique requirements and capabilities. Consequently, our analysis emphasises the importance of ongoing evaluation and iteration of CCMMs to ensure they remain effective tools in enhancing organisational cybersecurity postures.

\subsection{RQ1: Cybersecurity Capability Maturity Models (CCMMs)} \label{CCMMs}

Emerging from the thematic analysis \cite{Braun2012} of the literature, the classification of CCMMs presented in Figure \ref{fig:Results} categorises a diverse range of CCMMs according to their general applicability and targeted utility in specific sectors or distinct areas of cybersecurity such as cloud security or software security. This categorisation of CCMMs branches into 'Generic Models' and 'Specific Models,' with further subdivisions catering to the intricate aspects of their respective domains. 

Four of the included publications proposed non-specific maturity models: C2M2 \cite{C2M2}, CIP-based CCMM \cite{Dube2020}, Adaptive CCMM \cite{Ghaffari2018}, and \( (\text{CM})^2 \) \cite{Barclay2014} that serve as frameworks capable of being applied to a wide range of organisations, irrespective of their industry. C2M2, developed by the U.S. Department of Energy, is widely recognised in the literature for its comprehensive approach to cybersecurity assessments across both IT (Information Technology) and OT (Operational Technology) environments, applicable to organisations of all sizes and types \cite{Daraghmeh2021, Rea2017, Muronga2019, Curtis2015, Glantz2016}. Dube et al. \cite{Dube2020} have acknowledged the operational adaptability of the CIP-based CCMM in multiple sectors while primarily focusing on continuous improvement. Complementing these models, the Adaptive CCMM is highlighted for its flexible, level-based approach, tailored to enhance cybersecurity planning and improvement \cite{Ghaffari2018}. Meanwhile, designed to guide countries and organisations in attaining and maintaining a sustainable security advantage, \( (\text{CM})^2 \) is based on a 5-factor model that encourages the development of capabilities across various pillars, including society, technical, operational, business, legal, regulatory, and education/capability building measures \cite{Barclay2014}. These models have been designed to provide a universal set of guidelines that support the development of a robust cybersecurity posture across various operational fields \cite{Le2016}. 

The classification becomes more granular as the focus shifts to the more tailored Specific Models. The 'Critical Infrastructure' category encapsulates a variety of models that cater to different critical infrastructure sectors (i.e. energy, healthcare, transport, finance), highlighting the necessity for specialised approaches due to the unique vulnerabilities and stringent regulatory requirements of these industries \cite{Daraghmeh2021, Adler2013, Rojas2022, Brown2018}. ONG-C2M2 and ES-C2M2 as extensions of the foundational C2M2 framework place more emphasis on cybersecurity practices related to Operational Technology (OT) security, refinery and grid-specific threats, compliance with standards such as API\footnote{American Petroleum Institute} and NERC CIP\footnote{North American Electric Reliability Corporation Critical Infrastructure Protection}, and collaboration within the sectors \cite{ONGC2M2, ESC2M2, Walter2014, Le2016}. The FTCyberSec model proposed by Adler, extending the capabilities of ES-C2M2, brings a dynamic and iterative approach to cybersecurity management by simulating and analysing potential improvement strategies to enable organisations to formulate, test, and refine their plans for improving cybersecurity maturity levels \cite{Adler2013}. 

For the healthcare sector, the distinction is made between models focusing on cloud security, like the \( \text{M}^2CHS \) \cite{Akinsanya2019}, and those concentrating on data security, such as the EHR-MM and HDP-CMM, recognising the sector's increasing dependence on digital data storage and the critical need to safeguard patient information \cite{Daraghmeh2021, Rojas2022}. Expanding the scope of cybersecurity models, Alayo et al. \cite{Alayo2021} introduced the PFCMM, a model that integrates cybersecurity, cloud security, and privacy capabilities tailored explicitly for financial organisations in Peru. Similarly, Kour et al. expanded the C2M2 framework into the R-C2M2 for the railway industry, enhancing it with advanced security analytics and threat intelligence practices to meet sector-specific needs \cite{Kour2020}. Further contributing to this domain-specific approach, models like ICS-CTM2 and PSRC have been developed as easy-to-use tools, enabling organisations in the Industrial Control Systems sector to effectively evaluate and enhance their cybersecurity maturity \cite{Kreppein2021, Bhattacharya2022}. 

As revealed through our analysis, there is a range of maturity models within the critical infrastructure domain that is not confined to any specific industry but adaptable to various sectors within this domain \cite{Walter2014, Buzdugan2022}. For example, the AUMMCS-1 and Q-C2M2 models have been developed with specific regional focuses: AUMMCS-1 for the African Union Convention and Q-C2M2 for Qatar’s Cybersecurity Framework. Both models aim to improve the cybersecurity capabilities of organisations operating in critical infrastructure \cite{Von2015, Brown2018}. Solms \cite{Von2015} emphasises enhancing cybersecurity culture, fostering public-private partnerships, and prioritising education and training in the AUMMCS-1 model. In contrast, Brown \cite{Brown2018} focuses on aligning the Q-C2M2 model with Qatari legislation, ensuring legal compliance.

The Vulnerability-Driven C2M2, developed by Karabacak et al., adopts a more generalised approach to defining cybersecurity capabilities for organisations operating in the critical infrastructure sector \cite{Karabacak2016}. It establishes maturity criteria based on the root causes of vulnerabilities in critical infrastructures, making it applicable to a broader range of global cyber threats \cite{Karabacak2016}. 

Further classifications under the specific models include models for start-ups, communities and the workforce. The End-To-End CCMM, introduced by Selamat et al., allows start-ups to scale their security investments efficiently, adapting to the different stages of their lifecycle \cite{Selamat2022}. It simplifies the process of integrating cybersecurity measures in a way that is both effective and sustainable for emerging businesses \cite{Selamat2022}. As a community-focused model, CCSMM aims to foster a collaborative approach to cybersecurity within localities and smaller entities, promoting resource sharing and collective defence mechanisms \cite{White2011, Karabacak2016}. The Workforce category, with the NICE-C2M2, highlights the importance of human factors in cybersecurity, emphasising the development and assessment of the skills and knowledge of individuals who are responsible for maintaining the cyber health of organisations \cite{Walter2014, Rea2017, Le2016}.

Software Security and Cloud Security are two domains rapidly evolving with technology advancements \cite{Wang2018}. Models like the SD2-C2M2 and CSCMM are developed to address the unique challenges in these areas, ensuring that software development lifecycle and cloud-based operations maintain robust security postures against evolving threats \cite{Gourisetti2019, Le2017}.

The above bifurcation of literature into 'Generic' and 'Specific' categories helps understand these maturity models' broad applicability. While generic CCMMs offer a flexible approach that can be adapted across various industries \cite{Le2016}, the specific CCMMs deliver tailored guidance that aligns with the unique requirements of different industry sectors or different areas of security, such as cloud security and data security \cite{Sharkov2020}. Most of the specific models cater to assessing and enhancing cybersecurity capabilities in critical infrastructure sectors (i.e. energy, healthcare, transport, finance) \cite{ONGC2M2, ESC2M2, Kour2020, Adler2013, Dube2020, Kreppein2021, Drivas2020, Daraghmeh2021}. A few models, such as CSCMM, \( \text{M}^2CHS \) and HDP-CMM, have focused on particular aspects of security such as cloud security and data security, offering a structured set of practices for securing cloud environments, emphasising the need for robust access controls, data encryption, and ensuring the integrity, availability, and confidentiality of data \cite{Le2017, Akinsanya2019, Rojas2022}. The exploration of these models within the literature highlighted the strategic diversity in approaching cybersecurity. It also emphasised the importance of context-driven frameworks in achieving a robust and resilient cybersecurity posture capable of addressing the unique challenges presented by various security domains.

\subsubsection{{\textbf{Analysis of the Capability Domains in CCMMs:}}}\label{DomainAnalysis}
In order to delve deeper into the CCMMs, a thorough analysis of the capability domains within these models was conducted, as detailed in Table \ref{tab:ComparisonOfDomains}. This analysis aimed to evaluate the comprehensiveness of these models further and to explore how they differ, overlap, and complement one another across various models. A domain in this context refers to a specific area of cybersecurity management and operations, encompassing aspects like risk management, identity and access management, and incident response \cite{C2M2}. Each domain comprises a structured set of cybersecurity practices dedicated to that particular area of focus \cite{AlMatari2021}. Hence, the 20 domains outlined in Table \ref{tab:ComparisonOfDomains} serve as thematic pillars, each representing a distinct focus area that collectively constitutes an organisation's full spectrum of cybersecurity considerations \cite{Yigit2021}. 

When considering the generic models in Table \ref{tab:ComparisonOfDomains}, it is evident that out of those models, the Adaptive CCMM covers a wide range of domains, including risk management, identity and access management, threat and vulnerability management, event and incident response, workforce, personnel, and human resource management, and cybersecurity architecture and system security which are also covered in C2M2 \cite{C2M2, Ghaffari2018}. Adaptive CCMM includes unique domains such as technology and infrastructure, detection, monitoring and analysis, and physical security, demonstrating its comprehensive approach to addressing both digital and physical aspects of cybersecurity \cite{Ghaffari2018}.

Despite the Adaptive CCMM’s comprehensive approach \cite{Ghaffari2018} and the C2M2’s industry-vetted credentials \cite{C2M2}, both models exhibit specific gaps in their domain coverage that could significantly impact organisations. The absence of supply chain and external dependencies management in Adaptive CCMM \cite{Ghaffari2018} is a critical oversight, especially in today’s interconnected digital landscape, where third-party vendors and suppliers are pivotal in many industries \cite{Kreppein2021}. This gap could leave organisations vulnerable to risks from their supply chain, which can have cascading effects on their cybersecurity posture.

The lack of focus on education, training, and awareness in both Adaptive CCMM and C2M2 is a significant deficiency, particularly in an era of rapidly evolving cyber threats. Continuous education and heightened awareness are crucial for ensuring that staff at all levels are vigilant and well-informed about potential cybersecurity threats. This gap in the model could result in a workforce inadequately prepared to identify and respond to cyber threats, increasing the organisation’s vulnerability to cyber-attacks. Moreover, the absence of domains focused on communication, collaboration, and information sharing further exacerbates this issue, potentially hindering the effective exchange of crucial cybersecurity information and best practices. This limitation is particularly concerning as such collaboration is essential not only for individual organisations but also for fostering a resilient cybersecurity community capable of staying ahead of emerging threats.

The Adaptive CCMMs’ and C2M2s’ oversight in addressing legal and regulatory compliance poses a significant risk for organisations, potentially leading to non-compliance with relevant laws and industry standards, with ensuing legal consequences and reputational damage. This gap is compounded by the models’ lack of focus on aligning cybersecurity strategies with the business environment and strategy, which can result in misaligned priorities and inefficient resource allocation. A critical shortfall is the absence of a dedicated domain for business continuity and disaster recovery. In the wake of a cyber incident, the ability to rapidly recover and maintain operational continuity is crucial; without this focus, organisations are at risk of prolonged downtime and consequent financial and reputational damages. Furthermore, the model’s lack of emphasis on performance evaluation and improvement hinders an organisation’s ability to assess and enhance its cybersecurity measures effectively. In the constantly evolving landscape of cybersecurity, the necessity for continuous evaluation and iterative improvement is paramount, and the absence of this domain in the model could significantly impede an organisation’s ability to adapt and respond to new cyber threats.  

Similar gaps in domain coverage are also evident in specialised models like ES-C2M2, ONG-C2M2, FTCyberSec, R-C2M2, HDP-CMM, and B-C2M2 which have been adapted from the C2M2 framework for use in energy, transport, healthcare and build control systems sectors \cite{ESC2M2, ONGC2M2, Kour2020, Rojas2022, Glantz2016}. While tailored to address the unique requirements of their respective industries, these sector-specific models inherit the same gaps in domains as their parent model, C2M2.

\(\text{M}^2CHS\), EHR-MM, ICS-CTM2, PSRC, Vulnerability Driven CCMM, and AUMMCS-1, exhibit a narrowed concentration that overlooks many cybersecurity domains, including Data Privacy and Security, Incident Detection, Monitoring and Analysis, Legal, Regulatory and Compliance, Business Environment and Strategy, Business Continuity and Disaster Recovery, Physical Security and Performance Evaluation and Improvement domains \cite{ONGC2M2, ESC2M2, Kour2020, Akinsanya2019}. Furthermore, except PSRC, these models extend this pattern of omission to include crucial areas like Asset, Change, Configuration Management, Identity and Access Management, Threat and Vulnerability Management, and Event and Incident Response, which are more commonly found in previously discussed maturity models. Organisations following these models may find themselves ill-equipped to handle sophisticated cyber threats and face legal repercussions for non-compliance, leading to weakened defences against internal and external threats, inefficient management of cybersecurity resources, and a slower, less coordinated response to incidents, which collectively compromise their overall security and operational resilience \cite{Malatji2022}.

\begin{table}
\centering
\caption{Analysis of the Domains in CCMMs. Models are denoted using M1 to M23. M1-C2M2, M2-CIP based CCMM, M3-Adaptive CCMM, M4-\( (\text{CM})^2 \), M5-ONG C2M2, M6-ES C2M2, M7-FTCyberSec, M8-\( \text{M}^2CHS \), M9-EHR MM, M10-HDP MM, M11-PFCMM, M12-R C2M2, M13-ICS CTM2. '\checkmark' indicates that the capability domain is present in the model.}
\label{tab:ComparisonOfDomains}
\small{
\begin{tabular}{p{18em} c c c c c c c c c c c c c}
\toprule
\multirow{2}{*}{\textbf{Capability Domain}} & \multicolumn{4}{c}{\textbf{Generic Models}} & \multicolumn{7}{c}{\textbf{Specific Models}} \\
\cmidrule(rl){2-5} \cmidrule(rl){6-14}
 & M1 & M2 & M3 & M4 & M5 & M6 & M7 & M8 & M9 & M10 & M11 & M12 & M13 \\
\midrule
\rowcolor{lavender}
Risk Management & \checkmark & \null & \checkmark & \null & \checkmark & \checkmark & \checkmark & \checkmark & \checkmark & \checkmark & \checkmark & \checkmark & \null \\
Asset, Change and Configuration Management & \checkmark & \null & \checkmark & \null & \checkmark & \checkmark & \checkmark & \null & \checkmark & \checkmark & \checkmark & \checkmark & \checkmark  \\
\rowcolor{lavender}
Identity and Access Management & \checkmark & \null & \checkmark & \null & \checkmark & \checkmark & \checkmark & \checkmark & \checkmark & \checkmark & \checkmark & \checkmark & \null\\
Threat and Vulnerability Management & \checkmark & \null & \checkmark & \checkmark & \checkmark & \checkmark & \checkmark & \null & \null & \checkmark & \null & \checkmark & \null\\
\rowcolor{lavender}
Situational Awareness & \checkmark & \null & \null & \null & \checkmark & \checkmark & \checkmark & \null & \null & \checkmark & \null & \checkmark & \null\\
Event and Incident Response & \checkmark & \null & \checkmark & \null & \checkmark & \checkmark & \checkmark & \checkmark & \checkmark & \checkmark & \checkmark & \checkmark & \null\\
\rowcolor{lavender}
Supply Chain and External Dependency Management & \checkmark & \null & \null & \null & \checkmark & \checkmark & \checkmark & \null & \null & \checkmark & \null & \checkmark & \null\\
Cybersecurity Architecture & \checkmark & \checkmark & \checkmark & \null & \null & \null & \null & \null & \null & \checkmark & \checkmark & \null & \null\\
\rowcolor{lavender}
Cybersecurity Program and Policy & \checkmark & \checkmark & \checkmark & \null & \checkmark & \checkmark & \checkmark & \null & \checkmark & \checkmark & \null & \checkmark & \null\\
Workforce Management & \checkmark & \null & \checkmark & \null & \checkmark & \checkmark & \checkmark & \checkmark & \null & \checkmark & \null & \checkmark & \null\\
\rowcolor{lavender}
Education, Training and Awareness & \null & \checkmark & \checkmark & \checkmark & \null & \null & \null & \null & \checkmark & \null & \null & \null & \checkmark\\
Communication, Collaboration and Information Sharing & \null & \null & \null & \null & \checkmark & \checkmark & \checkmark & \null & \null & \null & \checkmark & \checkmark & \null\\
\rowcolor{lavender}
Data Security & \null & \null & \null & \null & \null & \null & \null & \null & \checkmark & \null & \checkmark & \null & \null\\
Technology and Infrastructure & \null & \null & \checkmark & \checkmark & \null & \null & \null & \null & \null & \null & \checkmark & \null & \checkmark\\
\rowcolor{lavender}
Incident Detection, Monitoring and Analysis & \null & \checkmark & \null & \null & \null & \null & \null & \null & \checkmark & \null & \checkmark & \null & \null\\
Legal and Regulatory Compliance & \null & \null & \checkmark & \checkmark & \null & \null & \null & \null & \null & \null & \checkmark & \null & \checkmark\\
\rowcolor{lavender}
Business Environment and Strategy & \null & \null & \null & \checkmark & \null & \null & \null & \null & \null & \null & \checkmark & \null & \null\\
Business Continuity and Disaster Recovery & \null & \null & \checkmark & \null & \null & \null & \null & \null & \checkmark & \null & \checkmark & \checkmark & \null\\
\rowcolor{lavender}
Physical Security & \null & \null & \checkmark & \null & \null & \null & \null & \checkmark & \null & \null & \null & \null & \null\\
Performance Evaluation and Improvement & \null & \null & \null & \null & \null & \null & \null & \null & \null & \null & \null & \null & \null\\
\bottomrule
\end{tabular}
}
\end{table}

Of the models catering to the critical infrastructure sector, Q-C2M2 stands out for its comprehensive coverage across a spectrum of domains, including risk management, incident response, and supply chain management \cite{Brown2018}. It emphasises workforce management and training, acknowledging the role of human factors in cybersecurity. However, the absence of Identity and Access Management raises concerns about effective access control to sensitive areas, while the lack of Threat and Vulnerability Management might leave vulnerabilities unaddressed. Similarly, insufficient focus on Situational Awareness could hinder proactive threat detection and response, and overlooking Physical Security overlooks the integral link between cyber and physical realms in infrastructure protection. 

\begin{table}
\centering
\caption*{Table \ref{tab:ComparisonOfDomains} (Continued). M14-PSRC, M15-AUMMCS1, M16-Vulnerability Driven CCMM, M17-Q C2M2, M18-B C2M2, M19- End to End CCMM, M20-CCSMM, M21-NICE C2M2, M22-SD2-C2M2, M23- CSCMM. '\checkmark' indicates that the capability domain is present in the model.}
\small{
\begin{tabular}{p{20em} c c c c c c c c c c}
\toprule
\multirow{2}{*}{\textbf{Capability Domain}} & \multicolumn{10}{c}{\textbf{Specific Models}} \\
\cmidrule(rl){2-11}
 & M14 & M15 & M16 & M17 & M18 & M19 & M20 & M21 & M22 & M23 \\
 \midrule
\rowcolor{lavender}
Risk Management & \checkmark & \null & \checkmark & \checkmark & \checkmark & \null & \null & \null & \null & \null \\
Asset, Change and Configuration Management & \checkmark & \null & \null & \checkmark & \checkmark & \null & \null & \null & \checkmark & \null \\
\rowcolor{lavender}
Identity and Access Management & \checkmark & \null & \null & \null & \checkmark & \null & \null & \null & \null & \checkmark \\
Threat and Vulnerability Management & \checkmark & \null & \null & \null & \checkmark & \null & \null & \null & \null & \null \\
\rowcolor{lavender}
Situational Awareness & \checkmark & \null & \null & \null & \checkmark & \null & \null & \null & \null & \null \\
Event and Incident Response & \checkmark & \null & \null & \checkmark & \checkmark & \null & \null & \null & \null & \null \\
\rowcolor{lavender}
Supply Chain and External Dependency Management & \checkmark & \null & \null & \checkmark & \checkmark & \null & \null & \null & \null & \null \\
Cybersecurity Architecture & \null & \null & \null & \checkmark & \null & \checkmark & \null & \null & \null & \null \\
\rowcolor{lavender}
Cybersecurity Program and Policy & \null & \checkmark & \checkmark & \checkmark & \checkmark & \null & \checkmark & \checkmark & \null & \checkmark \\
Workforce Management & \null & \null & \null & \checkmark & \checkmark & \checkmark & \null & \checkmark & \null & \checkmark \\
\rowcolor{lavender}
Education, Training and Awareness & \null & \checkmark & \checkmark & \checkmark & \null & \checkmark & \checkmark & \checkmark & \null & \checkmark\\
Communication, Collaboration and Information Sharing & \checkmark & \checkmark & \checkmark & \null & \checkmark & \null & \checkmark & \null & \null & \null \\
\rowcolor{lavender}
Data Security & \null & \null & \null & \checkmark & \null & \null & \null & \null & \null & \checkmark \\
Technology and Infrastructure & \null & \null & \null & \null & \null & \null & \null & \checkmark & \checkmark & \checkmark \\
\rowcolor{lavender}
Incident Detection, Monitoring and Analysis & \null & \null & \null & \checkmark & \null & \null & \null & \null & \null & \null \\
\rowcolor{lavender}
Business Environment and Strategy & \null & \null & \null & \null & \null & \null & \null & \null & \null & \null \\
Business Continuity and Disaster Recovery & \null & \null & \null & \checkmark & \null & \null & \checkmark & \null & \null & \null \\
\rowcolor{lavender}
Physical Security & \null & \null & \null & \null & \null & \null & \null & \null & \null & \null \\
Performance Evaluation and Improvement & \null & \null & \checkmark & \checkmark & \null & \null & \null & \null & \null & \null \\
\bottomrule
\end{tabular}
}
\end{table}

End-to-end CMM, CCSMM and NICE-C2M2, despite their distinct focuses on technology start-ups, community-level needs, and workforce development, respectively, share a common shortfall: the absence of several key cybersecurity domains that are commonly present in other models. These missing domains include Risk Management, Identity and Access Management, Asset Management, Threat and Vulnerability Management, and Event and Incident Response. It limits the models' capability to proactively identify, assess, and mitigate potential cyber risks, fails to adequately safeguard sensitive information and assets, and might not be equipped to identify and address evolving cybersecurity threats effectively, thereby compromising the overall security posture \cite{Malatji2022}.

SD2-C2M2, with its emphasis on Asset, Change, Configuration Management, and Technology and Infrastructure, shows a strong orientation towards maintaining system integrity and managing technological assets effectively \cite{Gourisetti2019}. However, its lack of domains like Risk Management and Event and Incident Response suggests a potential vulnerability in proactively identifying and mitigating cyber threats and in responding to incidents efficiently. In contrast, CSCMM covers a broader range, including Identity and Access Management, Cybersecurity Programs, Policy and Governance, and Legal, Regulatory and Compliance, indicating an approach that incorporates both the technical and organisational aspects of ensuring cloud security in organisations \cite{Le2017}. This model, however, shares a similar gap with SD2-C2M2 in lacking domains such as Risk Management and Threat and Vulnerability Management, which are crucial for a comprehensive cybersecurity strategy. The absence of these key areas in both models could result in under-preparedness against evolving cyber threats and challenges in adapting to regulatory changes, potentially leaving organisations vulnerable to unaddressed security risks and compliance issues. Therefore, while CSCMM offers a more rounded approach compared to SD2-C2M2, both models could benefit significantly from incorporating the missing domains to enhance their effectiveness in addressing the challenges of cybersecurity.

In summary, the analysis underscores the importance of continuous evolution in CCMMs to ensure they address a comprehensive range of cybersecurity concerns. This includes not only technical and procedural aspects but also the human, communicative, and strategic elements crucial for an effective and resilient cybersecurity posture across various sectors \cite{Yigit2021}.

\subsection{RQ2: Limitations of CCMMs}\label{Limitations}

The analysis of the literature on the existing cybersecurity capability maturity model revealed a range of limitations that can undermine their effectiveness and applicability across diverse operational contexts. To explore these limitations distinctly, an analysis rooted in socio-technical systems theory \cite{Whitworth2009, Malatji2019} was employed. This theory is used to examine the constraints of current CCMMs based on the social, technical, and environmental factors  \cite{Sony2020} that can affect the efficacy of these models. The social dimension comprises human and organisational factors (i.e. organisational culture, user behaviour, and management practices), the technical dimension includes technology, tools and work activities, and the environmental dimension covers attributes that influence both the social and technical aspects such as regulatory requirements and industry standards \cite{Malatji2019}. Figure \ref{fig:Results} demonstrates the categorisation of the limitations in CCMMs into the above dimensions, providing a structured overview of the areas where CCMMs fall short. 

One of the critical shortcomings related to the social dimension is the insufficient emphasis on cybersecurity culture and awareness \cite{Barclay2014, Akinsanya2019, Muronga2019}. The importance of nurturing a security-conscious environment within an organisation cannot be overstated, as it is often the first line of defence against cyber threats \cite{Barclay2014}. The lack of practices in CCMMs ensuring cybersecurity culture might leave organisations vulnerable to risks that stem from human error or lack of awareness \cite{Muronga2019}.

The limited focus on human and behavioural factors, such as incorporating user-centric security practices and strategies to mitigate insider threats and enhance the overall security mindset of the organisation in these models, could lead to the development of cybersecurity strategies that do not fully account for the impact of human actions on organisational cybersecurity \cite{Buzdugan2022}. This oversight can result in the implementation of technical measures that are disconnected from the realities of human behaviour, reducing their overall effectiveness in mitigating cyber risks \cite{Buzdugan2023}.

Customisation of these models to specific organisational needs also poses a challenge \cite{Selamat2022}. The difficulty in tailoring these models to cater to the unique requirements of different organisations can impede their effectiveness across various contexts \cite{Drivas2020, Selamat2022, Akinsanya2019, Dube2020}. This lack of customisation can be particularly problematic in organisations with unique operational structures or those operating in specialized industries, where a one-size-fits-all approach to cybersecurity may not be sufficient \cite{Drivas2020, Dube2020}.

Drivas et al. \cite{Drivas2020} have discussed the lack of implementation feedback as another limitation of the existing models. Without facilitating a mechanism for collecting and analysing feedback on their implementation process and outcomes, organisations may not receive the necessary insights to continuously refine and improve their cybersecurity strategies \cite{Drivas2020}. This absence of feedback can impede the process of learning and adapting cybersecurity practices over time, which is crucial in the ever-evolving landscape of cyber threats.

In the technical dimension, one major limitation noted in the identified articles is the inherent complexity of the existing CCMMs \cite{Georgiev2021, Selamat2022, Brown2018, Le2016}. This complexity can pose a substantial barrier to understanding and implementing these models, particularly for organisations that lack extensive cybersecurity expertise \cite{Ghaffari2018}. Furthermore, many CCMMs do not adequately cater to organisations with multiple interdependent systems \cite{Drivas2020, AlMatari2021, Walter2014}. Such a shortfall can result in gaps in cybersecurity strategies, especially in sectors where system interdependencies are prevalent, such as critical infrastructure or large-scale enterprises \cite{Walter2014, Curtis2015}.

The models' lack of extensibility to emerging threats implies that organisations relying on these models might find themselves ill-equipped to adapt to new and evolving cyber threats \cite{Aliyu2020, Akinsanya2019, Drivas2020, Le2017, Opeoluwa2019}. The inability of these models to evolve with the changing cybersecurity landscape can significantly impede an organisation's capacity to respond effectively to future cybersecurity challenges \cite{Drivas2020, Opeoluwa2019}. 

Moreover, the lack of integration with existing cybersecurity frameworks represents a missed opportunity for synergy and comprehensive coverage \cite{Brown2018, Walter2014}. When CCMMs operate in isolation from other established cybersecurity frameworks and initiatives, it can lead to a disjointed and potentially less effective approach to cybersecurity \cite{Brown2018}. Integrating CCMMs with other frameworks can enhance their effectiveness by providing a more comprehensive, multi-faceted approach to cybersecurity that leverages the strengths of various frameworks and methodologies \cite{Brown2018, Walter2014}.

Another key issue is that some models are characterised as static and diagnostic, lacking direct actionability \cite{Drivas2020, Le2016, Adler2013}. This characterisation implies that these models might not offer clear, actionable steps for organisations looking to enhance their cybersecurity posture \cite{Le2016}. As a result, there could be a notable disconnect between what is assessed through these models and the tangible improvements that need to be made in practice \cite{Drivas2020}. Such a gap can hinder the effective translation of assessment insights into meaningful cybersecurity enhancements. 

There is often an overemphasis on compliance within these models \cite{Drivas2020, Le2016, Le2017, Opeoluwa2019} that can be categorized in this dimension. This focus can inadvertently lead organisations to adopt a checkbox mentality, where the primary goal becomes meeting specific compliance requirements rather than genuinely enhancing their cybersecurity posture \cite{Le2017}. While compliance is undoubtedly important, over-focusing on it can detract from the broader objective of developing robust and effective cybersecurity measures \cite{Le2016}. This overemphasis on compliance may result in a superficial approach to cybersecurity, where organisations do the minimum required to meet standards rather than striving for comprehensive security improvements \cite{Opeoluwa2019}.

Additionally, the limited applicability of some models across various organisational types and situations suggests that their utility may be restricted \cite{Rojas2022, Dube2020, Karabacak2016}. This limitation indicates that these models may not be versatile enough to be effectively implemented across a diverse range of organisational contexts, which can be a significant drawback for organisations with unique cybersecurity needs or those operating in specialized sectors \cite{Rojas2022}.

The lack of emphasis on the return on investment (ROI) associated with cybersecurity initiatives \cite{Selamat2022, Brown2018, Buzdugan2023} can also be taken into consideration under the environmental dimension. This gap means that organisations may struggle to quantify the financial benefits or effectiveness of their cybersecurity investment in implementing the cybersecurity capability maturity model, making it challenging to justify these expenditures or assess their impact on the overall security posture. Without a clear understanding of ROI, organisations might find it difficult to allocate resources optimally and make informed decisions about their cybersecurity strategies \cite{Brown2018}.

Lastly, the absence of a holistic perspective in many CCMMs is another limitation discussed in the literature \cite{Aliyu2020, Le2016, Opeoluwa2019, Alayo2021}. When the models are incapable of providing a complete view that encompasses all aspects of an organisation's cybersecurity maturity - often described as a '360 degree perspective' - they may fail to fully integrate cybersecurity considerations into every facet of organisational strategy and operations \cite{Opeoluwa2019}. A '360-degree perspective' refers to evaluating and enhancing cybersecurity capabilities across all organisational layers, including technical defences, management processes, governance structures, and employee behaviour \cite{Le2016}. Without such a holistic perspective, where cybersecurity maturity is assessed and developed as an integral part of every business function, organisations might end up with a disjointed cybersecurity posture. For example, a CCMM might emphasise improving technical controls while neglecting the maturity of governance or risk management processes, leading to gaps that undermine overall cybersecurity effectiveness \cite{Le2016}. A truly holistic CCMM ensures that cybersecurity capability maturity is not treated in isolation within specific areas but is comprehensively embedded across the entire organisational framework, ensuring alignment with the organisation's overall strategy and operational goals \cite{Aliyu2020}.

In summary, while existing CCMMs offer valuable frameworks for assessing and enhancing cybersecurity, these limitations highlight areas where these models could be further developed to provide more comprehensive, practical, and adaptable tools for organisations across different sectors and contexts. Addressing these gaps is crucial for ensuring that organisations can effectively navigate the complex and evolving landscape of cybersecurity.

\subsection{RQ3: Challenges}\label{Challenges}

While CCMMs offer a structured approach to improving an organisation's cyber resilience, their implementation and adoption can be hindered by a complex interplay of factors, as shown in Figure \ref{fig:Results}. This section explores the challenges organisations face, categorised using the TOE framework \cite{Awa2017, Wallace2021}. Through the thematic analysis, we identified key themes that fall under the categories of Technological, Organisational, and Environmental (TOE) influences. By examining these themes, we gain an understanding of the obstacles organisations must overcome to successfully leverage CCMMs for enhanced cybersecurity posture. 

\subsubsection{Challenges for Implementation}\label{ImplementationChallenges}\hfill

\hspace{-1.2em}
From a technological perspective, the inherent complexity of CCMMs emerges as a significant challenge during implementation. This can be particularly daunting for organisations lacking a robust cybersecurity foundation \cite{Walter2014}. Understanding and effectively deploying these models can be difficult due to their detailed nature \cite{Selamat2022}. Organisations, especially those with limited cybersecurity expertise, may struggle to navigate these complexities, potentially leading to misinterpretations and flawed implementations \cite{Walter2014}.  This can result in a mismatch between the models' recommendations and the organisation's actual cybersecurity needs, ultimately hindering the effectiveness of the CCMM implementation \cite{Baikloy2020}. Moreover, the integration of CCMMs with existing cybersecurity processes and frameworks can present technological challenges, such as compatibility issues with current security tools or difficulties in aligning the model’s recommendations with the organisation's technological infrastructure. Organisations often have already established security practices, and introducing a new CCMM can require significant adjustments, such as updating legacy systems, retraining staff on new tools and processes, or ensuring that the CCMM’s requirements do not conflict with existing technological standards and protocols \cite{Rea2017}. Seamless integration requires meticulous planning and execution to ensure the CCMM complements and strengthens existing processes, not disrupt them \cite{Brown2018}.  

From an organisational perspective, fostering individual responsibility and organisation-wide engagement presents significant challenges. Buzdugan et al. highlight that it is crucial to ensure every member understands and takes ownership of cybersecurity for successful implementation \cite{Buzdugan2022}. Without a strong cybersecurity culture where individuals feel accountable, implementation efforts may fall short \cite{Buzdugan2023}. Engaging the entire organisation is equally critical \cite{Buzdugan2022}. Failure to secure active participation and buy-in from all departments and levels can lead to CCMM being implemented in isolation without a unified organisational strategy \cite{AlMatari2021}. For example, if the IT department focuses only on technical controls and infrastructure security, without HR’s involvement in training and enforcing security policies and executive management’s engagement in prioritising cybersecurity to align with business objectives, the efforts in improving cybersecurity may be fragmented, leading to vulnerabilities where certain areas are well-protected while others are neglected. This undermines the model's effectiveness.

Additionally, resource constraints and skills gaps can hinder the effective implementation of CCMMs \cite{Brown2018, Dube2021}. Implementing CCMMs often necessitates substantial financial investment, skilled personnel, and dedicated time. However, organisations, especially smaller ones, may struggle with these demands. The overhead costs associated with implementation can be prohibitive, while the lack of in-house cybersecurity expertise can further complicate the process \cite{Walter2014, Brown2018}. Compounding the aforementioned challenges is the lack of clear implementation guidance. As Brown et al. points out, this absence leaves organisations without a roadmap for deployment \cite{Brown2018}. Identifying starting points and effectively integrating the CCMM with the organisation's existing cybersecurity framework becomes a significant challenge \cite{Walter2014}. 

The rapidly evolving regulatory landscape in cybersecurity presents a significant environmental challenge for CCMM implementation. Organisations need to continuously adapt their security posture to comply with both existing and emerging regulations, which can include internal organisational policies, government-mandated laws and industry-specific standards \cite{Aliyu2020, Alayo2021}. This necessitates flexible and up-to-date CCMMs, which can be difficult to maintain \cite{Alayo2021}. Without this flexibility, organisations risk falling out of compliance with new regulations, potentially facing legal and financial repercussions \cite{Aliyu2020}.

In conclusion, implementing CCMMs presents a complex landscape of challenges. These range from technological hurdles like model integration to fostering a strong cybersecurity culture within the organisation. Resource constraints and the rapidly evolving regulatory environment further complicate the process. Overcoming these challenges necessitates a tailored approach \cite{Daraghmeh2021, Le2017, Aliyu2020}. Organisations must consider their unique needs and the broader cybersecurity landscape to effectively utilize CCMMs. By addressing these complexities, organisations can unlock the potential of CCMMs and build a robust defence against evolving cyber threats.

\subsubsection{Challenges for Adoption}\label{AdoptionChallenges}\hfill

\hspace{-1.2em}

One of the significant technological challenges hindering the adoption of CCMMs is their perceived complexity \cite{Aliyu2020}. These models are often designed to be comprehensive, encompassing a wide range of cybersecurity best practices and control objectives. While it ensures a thorough assessment of an organisation's cybersecurity posture, it can also lead to a perception that CCMMs are intricate and difficult to understand \cite{Akinsanya2019}. It can create a sense of intimidation among organisations, particularly those with limited cybersecurity expertise \cite{Aliyu2020}. Without a clear understanding of the model, organisations may misapply its recommendations \cite{Walter2014}. Moreover, if employees find the model difficult to grasp, they may be less likely to actively participate in the adoption process \cite{Buzdugan2022}. 

CCMMs typically outline a well-defined set of control objectives and best practices. In contrast, existing processes may have evolved over time to meet specific organisational needs and may not perfectly align with the model's structure \cite{Aliyu2020, Walter2014}. This can lead to friction during adoption, as organisations struggle with how to accommodate existing workflows with the requirements of the CCMM \cite{Brown2018}. A poorly integrated CCMM can lead to inefficiencies, duplicated efforts, and confusion among employees \cite{Akinsanya2019}. Even if the model is initially adopted, these challenges can discourage its continued use, leading to a decline in adherence and eventually undermining the effectiveness of the CCMM in improving the organisation's cybersecurity posture. 

From a technological perspective, the inherent static nature of CCMMs also raises concerns about their long-term effectiveness \cite{Adler2013}. Organisations may be hesitant to adopt a model that may not fully address the dynamic nature of cyber threats \cite{Adler2013}. Additionally, organisations may become fixated on achieving a particular level within the model, neglecting the need for ongoing monitoring and adaptation of their cybersecurity posture \cite{Brown2018}. This focus on static goals can hinder the crucial practice of observability, the ability to continuously monitor and analyse security data to identify and respond to emerging threats \cite{Brown2018, Walter2014}.

Competing organisational priorities emerge as a challenge in adopting CCMMs. Organisations may not always perceive cybersecurity as a critical issue compared to immediate business needs \cite{Brown2018}. This can lead to the sidelining or under-resourcing of CCMM adoption. When cybersecurity initiatives compete for attention and resources, the adoption process can be significantly hindered \cite{Brown2018}. Also, the gaps in understanding and commitment across different levels of management can pose a significant barrier. These knowledge and commitment disparities can lead to a lack of clear leadership and unified direction for cybersecurity initiatives \cite{AlMatari2021}.  Without strong, cohesive management support, the adoption of CCMMs risks becoming fragmented, resulting in inconsistencies in cybersecurity strategies across the organisation \cite{Buzdugan2022}.

A critical but often overlooked challenge in adopting CCMMs lies in the need to initiate change at the individual level \cite{Buzdugan2022}. Many employees, particularly those outside of technical roles, may view cybersecurity as solely the domain of IT specialists \cite{Adler2013}. This perception can lead to a sense of disengagement and a belief that individual actions have little impact on the organisation's overall cybersecurity posture. 

Effectively navigating the CCMMs, interpreting their requirements, and integrating them with existing security structures requires an understanding of cybersecurity best practices and the specific model being adopted. Organisations lacking in-house expertise may need to rely on external consultants, adding to the overall cost \cite{Walter2014}. Furthermore, the costs associated with acquiring and maintaining the necessary software tools and training programs can add to the financial burden \cite{Brown2018}. Moreover, the adoption process itself requires a considerable effort from various departments across the organisation. Employees need to be actively involved in assessments, training programs, and potentially workflow adjustments \cite{Adler2013}. This can place a strain on existing resources and potentially disrupt day-to-day operations \cite{Adler2013}.

Adler et al. highlight that the absence of government mandates for implementing CCMMs presents a complex environmental challenge \cite{Adler2013}. While CCMMs offer a structured approach to enhancing cybersecurity posture, their adoption in many sectors remains voluntary. This lack of mandated adoption creates an uneven playing field where some organisations prioritise cybersecurity investments while others do not \cite{Adler2013, Brown2018}. Organisations that choose not to invest in adopting a CCMM may benefit from the improved cybersecurity posture of their competitors who have invested in these models \cite{Brown2018}. This can create a disincentive for widespread adoption, hindering the overall improvement of cybersecurity within the broader ecosystem. 

In conclusion, the path towards adopting CCMMs is not without its complexities. These challenges encompass technological hurdles like model complexity and integration, organisational barriers such as competing priorities and cultural gaps, and environmental considerations, including the evolving regulatory landscape.  Successfully navigating this landscape requires a tailored approach. Organisations must carefully assess their unique needs and resources while remaining mindful of the broader cybersecurity environment. By acknowledging these challenges and developing a comprehensive strategy to address them, organisations can utilise the full potential of CCMMs.

\section{Discussion}
\label{Section5-Discussion}

In this systematic literature review, we analysed 43 publications, adhering to the guidelines proposed by Kitchenham and Charters \cite{Kitchenham02} and aligning our approach with the PRISMA framework. We identified 23 distinct CCMMs that served various organisations' cybersecurity needs, addressing various challenges and objectives. To answer research question RQ1, we analyzed the domains covered by these maturity models to understand their scope and the various aspects they encompass. Additionally, we explored the inherent limitations of these models and the challenges organizations may face in implementing and adopting them, which informed our responses to research questions RQ2 and RQ3.

In the discussion that follows, we provide a detailed examination of these findings. We first expand on the domains covered by CCMMs, highlighting how they address different facets of cybersecurity across organisations. Next, we strategize beyond the identified limitations, proposing approaches to enhance the effectiveness of CCMMs. Finally, we address the implementation and adoption challenges, offering insights into how organizations can navigate these issues to successfully deploy and utilize CCMMs. 

\subsection{Expanding CCMM Domains}

As discussed in Section \ref{CCMMs}, the literature has highlighted 'Generic' and 'Specific' CCMMs where each type addresses distinct needs. Generic models tackle a broad spectrum of general cybersecurity requirements, while Specific models target the unique challenges of particular sectors or security concerns \cite{Rabii2020, Buzdugan2023}. However, despite their strengths, both types of CCMMs can exhibit common shortcomings, particularly in the domains they cover.

While Generic models like C2M2 and Adaptive CCMM provide broad coverage, they lack depth in crucial areas. Integrating domains like Training and Awareness, Communication and Collaboration, Legal Compliance, and Physical Security is essential for building comprehensive security strategies. By fostering a culture of security awareness through employee training, organisations can empower their workforce to identify and report suspicious activity \cite{Gourisetti2020}. This proactive approach significantly reduces the attack surface and minimises the potential for successful cyberattacks \cite{Ghaffari2018}. Effective communication and collaboration within the organisation is paramount for a robust security posture. It facilitates clear communication channels and collaboration with other departments like IT and HR, and the organisation can take swift and coordinated action to investigate and contain potential threats \cite{White2011}. 

The legal landscape adds another layer of complexity. Operating within a web of regulations, organisations must ensure alignment with legal and regulatory requirements, particularly those related to data privacy and security. Compliance minimises the risk of hefty penalties, and fosters trust with customers and partners \cite{Brown2018}. Additionally, the intricate link between cybersecurity and physical security cannot be ignored. A cyberattack, for instance, might not only compromise digital systems but also disrupt physical operations by manipulating critical infrastructure controls. Integrating physical security considerations into CCMMs ensures a more holistic approach to cybersecurity, safeguarding critical assets from multifaceted attacks \cite{Alayo2021, Kour2020}. This can prevent physical damage, operational disruptions, and potential safety hazards \cite{Brown2018}.

Similarly, the Adaptive CCMM could benefit from the inclusion of Situational Awareness and IDA (Incident Detection and Analysis) domains. Situational Awareness involves continuous analysis and interpretation of security data to understand the current state of the organisation's security posture \cite{Rojas2022}. This allows for proactive threat detection. By identifying unusual activity, organisations can investigate and potentially neutralise a threat before it can compromise systems or steal data \cite{Kreppein2021}. On the other hand, IDA goes beyond simply identifying security incidents. It equips organisations to conduct a thorough investigation, understand the root cause of the incident, and implement effective mitigation strategies \cite{Brown2018}. This not only minimises damage but also provides valuable insights for future improvements \cite{Daraghmeh2021}.

Models with a narrower focus, like CIP-based CCMM and \((\text{CM})^2\), have overlooked key areas such as Risk Management, Asset Management, Identity and Access Management, Incident Response, External Dependencies Management, and Communication and Collaboration. The risk management domain fosters proactive identification, prioritisation, and mitigation of security risks, empowering organisations to allocate resources strategically. A robust risk management program would employ practices like threat modelling and vulnerability assessments to identify vulnerabilities before attackers exploit them \cite{C2M2}. Another critical missing domain is Asset Management. Maintaining a complete inventory of all assets, including hardware, software, and data, ensures effective protection and control \cite{Ghaffari2018}. Strong asset management practices involve activities like asset discovery and classification \cite{Bhattacharya2022}. If asset management is neglected, organisations remain vulnerable to potential attacks targeting unidentified assets. Discovery and classification practices would identify the assets in the organisation, allowing them to implement appropriate security controls - both technical and non-technical - including educational and training interventions.

Furthermore, rigorous control over access to critical systems and data is essential. A well-defined Identity and Access Management (IAM) domain minimises unauthorised access and potential breaches by implementing a series of best practices \cite{Kreppein2021, Le2017}. The practices include granting and revoking access and ensuring that only authorised personnel have access to critical systems and data \cite{Rojas2022}. This minimises the risk of unauthorised access even if credentials are compromised. The IAM domain also facilitates continuous monitoring and auditing of user activity and allows for the identification of suspicious behaviour \cite{C2M2}. This enables organisations to detect and respond to potential security incidents before they can escalate.

Incorporating an Incident Response domain enables organisations to respond to security incidents swiftly and effectively. This domain covers clear procedures for identifying, mitigating, and recovering from security breaches \cite{C2M2}. This minimises downtime and data loss. Practices such as incident detection and containment procedures empower organisations to isolate the infected systems, prevent further data or asset compromisation, and restore operations efficiently \cite{Glantz2016}.

The security posture of an organisation is only as strong as its weakest link \cite{Kreppein2021}. Many organisations rely on third-party vendors for critical services \cite{Kreppein2021}. Therefore, having an External Dependencies Management domain in the maturity model adopted by the organisation helps to assess and mitigate security risks posed by these external dependencies proactively. External Dependency Management practices include conducting vendor security audits and incorporating security clauses into contracts \cite{Rojas2022}. By proactively assessing the security posture of the vendors, an organisation can identify potential risks and take steps to mitigate them, minimising the impact of a third-party breach \cite{Rojas2022}.

The incorporation of a Performance Evaluation and Improvement domain in CCMMs is notably absent across the board. This gap is significant, as this domain encompasses critical practices such as regular cybersecurity assessments, feedback loops for process improvement, and the use of quantifiable metrics to track improvements over time \cite{Buzdugan2023}. Regular assessments enable an organisation to continuously monitor and evaluate its cybersecurity measures, identifying vulnerabilities proactively. However, without this domain, CCMMs leave organisations without a systematic approach to validate the effectiveness of their cybersecurity strategies. Feedback loops are essential for learning from incidents and assessments, fostering a culture of continuous improvement \cite{Rea2017}. The use of quantifiable metrics to measure improvement offers a concrete means of tracking cybersecurity progress. Without this, organisations are at a disadvantage, lacking tangible evidence to inform decision-making and validate the impact of their cybersecurity initiatives \cite{Dube2020}.

The inclusion of specific domains such as data security, network security, endpoint security, and cloud security within CCMMs is also paramount for a holistic cybersecurity strategy, as organisations today face a complex cybersecurity landscape with threats targeting various aspects of their IT infrastructure and data \cite{Sharkov2020}. The data security domain focuses on protecting the confidentiality, integrity, and availability of sensitive data \cite{Bahuguna2019}. Practices like data classification and encryption ensure data remains protected even in the event of a breach \cite{Divadari2023}. Similarly, the network security domain safeguards the organisation's network infrastructure. Firewalls, intrusion detection systems, and vulnerability management practices \cite{Srivastava2024} within this domain monitor network traffic for malicious activity, prevent unauthorised access and identify and patch vulnerabilities before attackers can exploit them \cite{Barclay2014}.

Moving to individual devices, the endpoint security domain focuses on protecting laptops, desktops, and mobile phones used by employees \cite{Mishra2022}. Practices like endpoint protection software, application whitelisting, and patch management ensure these devices are secure and don't introduce vulnerabilities into the network \cite{Mishra2022}. Furthermore, with the increasing adoption of cloud computing, securing cloud-based resources is critical. The cloud security domain emphasises practices like secure cloud configurations, data encryption in transit and at rest, and robust access controls for cloud environments \cite{Le2017}.

A truly comprehensive CCMM also requires a Business Continuity and Disaster Recovery (BC/DR) domain, ensuring that critical business functions can be restored quickly and efficiently in the event of a disruptive event \cite{Ghaffari2018}. Disruptions can take many forms, including natural disasters, power outages, or human error. Without a BC/DR plan in place, the organisation may struggle to resume normal business functions in the event of a disruption, potentially leading to significant financial losses and reputation damage \cite{Alayo2021}. The BC/DR domain incorporates impact assessments that regularly evaluate the potential impact of different disruptions on critical business functions \cite{White2011}. It allows organisations to prioritise resources and identify essential systems and data that require the highest level of protection \cite{Ghaffari2018}. Another vital practice is the development of business continuity plans \cite{Altaha2023}. These detailed plans define alternative processes and resources that can be used to maintain critical business functions during a disruption \cite{Brown2018}. By fostering these practices, the BC/DR domain ensures that cybersecurity is not conducted in isolation \cite{White2011}. 

As none of the existing models includes a domain dedicated to fostering a cybersecurity culture \cite{Barclay2014, Akinsanya2019}, we also highlight the importance of having a Cybersecurity Culture domain within a CCMM to ingrain security in everyday practices and decision-making \cite{Georgiadou2022}. It encourages the promotion of security champions within departments that can act as liaisons between employees and IT security professionals, fostering a collaborative environment and promoting best practices amongst peers \cite{Georgiadou2022}. This domain also promotes practices like integrating security considerations into performance reviews \cite{Gcaza2017}. Linking security awareness and adherence to security policies with employee evaluations encourages a sense of ownership and accountability for maintaining a strong security posture \cite{Gcaza2017}. Furthermore, the domain encourages the use of internal knowledge-sharing platforms. These platforms allow employees to share best practices and insights gleaned from security training or personal experiences, fostering a collective learning environment within the organisation \cite{Georgiadou2022}.

The gaps in the domains identified in existing CCMMs highlight the need for a more comprehensive approach. By incorporating additional domains, CCMMs can evolve to empower organisations to build a holistic and adaptable security posture. This will equip organisations to not only address current threats but also proactively prepare for the evolving cybersecurity threat landscape.

\subsection{Strategizing Beyond Limitations}

The inherent limitations of CCMMs highlight the need for strategic improvements. By addressing these shortcomings, we can create more robust and adaptable CCMMs. We propose the following strategies to overcome these limitations, ensuring CCMMs evolve into powerful tools that enhance cybersecurity across a wide range of operational contexts. 

\subsubsection{Tailored framework for diverse needs.}
The rigid nature of existing CCMMs struggles to accommodate the unique requirements and complexities of diverse organisational structures \cite{Drivas2020}. Recognising that organisations vary greatly in their operational contexts, risk profiles, and resource availability, a tailored cybersecurity capability maturity framework can be developed by incorporating customizable domains and practices that can be adjusted according to organisational needs. This customization allows organisations to selectively focus on capability domains and practices that are most pertinent to their specific security concerns, operational realities, and strategic goals \cite{Hauck2011}. By enabling such tailored implementation paths, organisations are not only able to align their cybersecurity efforts with their available resources but can also prioritise initiatives that address their most pressing risk areas \cite{Spruit2014}. This method offers a dynamic solution that transcends the one-size-fits-all approach, ensuring that CCMMs can be effectively applied in varied settings—from small enterprises with limited cybersecurity budgets to large corporations facing intricate cyber threats due to their vast, interconnected systems \cite{Bahuguna2019}. Furthermore, the adaptability provided by this strategy ensures that the cybersecurity framework remains both relevant and robust across different sectors. Organisations benefit from a CCMM that is aligned with their immediate needs and capable of evolving in response to changes in their operational environment or threat landscape. 

\subsubsection{Expanding domains and practices for a dynamic and comprehensive cybersecurity approach.} 
Integrating a spectrum of domains encompassing technical, organisational and human aspects related to cybersecurity into the maturity framework ensures a holistic perspective on cybersecurity maturity \cite{Awa2017}. This inclusivity allows for the cultivation of a pervasive cybersecurity culture that raises awareness across all levels of the organisation and embeds security consciousness into actions and decisions \cite{Jeong2022}. It also brings human and behavioural factors into the forefront of cybersecurity strategies, acknowledging the critical role that individuals play in both the creation and prevention of security risks \cite{Barclay2014}.

To ensure a cohesive and consistently applied cybersecurity strategy across all organisational levels, integrating specific practices within the governance and operational domains is crucial. This includes establishing regular cross-functional cybersecurity meetings and creating liaison roles to facilitate continuous communication between governance bodies and operational teams \cite{White2011}. Utilising shared cybersecurity management platforms and collaborative documentation tools enhances visibility and ensures everyone is aligned on the organisation’s cybersecurity posture and initiatives \cite{Von2015}. 

Adopting practices that actively engage with and mitigate real-world threats and vulnerabilities is essential to transcend mere compliance and cultivate genuine security within organisations. Integrating threat intelligence to proactively identify emerging risks, conducting detailed risk assessments beyond standard compliance to prioritise threats based on potential impact, and implementing tailored security awareness training can significantly elevate an organisation's defence mechanisms \cite{Schlette2021, Kulugh2022}. Additionally, employing continuous monitoring for real-time threat detection, coupled with robust incident response plans, ensures swift action against security breaches. Regular testing of security control efficacy through penetration testing and vulnerability scanning, along with establishing feedback mechanisms for continuous security enhancement from all organisational levels, reinforces a proactive security posture \cite{Shaked2020}. Moreover, conducting thorough post-incident analyses to extract lessons learned from specific breaches or security incidents further solidifies the organisation's resilience against future threats \cite{Baikloy2020}. These comprehensive practices foster a culture of continuous risk management and improvement, ensuring organisations achieve a security stance that is not only compliant but genuinely secure and adaptive to the dynamic cybersecurity landscape.

\subsubsection{Metrics-driven cybersecurity enhancement.} 
Embedding quantitative and qualitative metrics with the cybersecurity practices in the capability domains enables measuring the impact of the organisation's cybersecurity efforts and understanding the financial benefits of their security investments \cite{Karabacak2016}. As the existing CCMMs fall short of offering a clear, data-driven picture of how security measures contribute to organisational objectives, adding metrics and Key Performance Indicators (KPIs) into the models allows organisations to assess the strength of their security posture and articulate and demonstrate the value of cybersecurity investments to stakeholders \cite{Adler2013}. The adoption of a metrics-driven approach aligns cybersecurity initiatives with broader organisational goals, ensuring that resources are allocated efficiently and strategically. Furthermore, the emphasis on data provides a solid foundation for continuous refinement of security measures, enabling organisations to adapt proactively to evolving threats and changing business needs \cite{Akinsanya2019, Le2017}.

\subsubsection{User-centered cybersecurity tools and processes.} 
The complexity of the existing CCMMs can be overcome by translating CCMM components into intuitive, user-centred tools and guidelines \cite{Adler2013}. Such a tool makes the maturity assessment process accessible for organisational entities with varying levels of expertise, particularly small to medium-sized enterprises \cite{Adler2013, Le2017}. When tools are designed with the user in mind (i.e., through the user-centred approaches ) \cite{Reynolds2021, Koh2011}, 
they enhance the interaction between the user and the CCMM assessment process by breaking down complex maturity model components into simpler, more understandable tasks. They provide clear, step-by-step instructions, helping users accurately assess their organisation’s cybersecurity maturity without being overwhelmed by technical jargon or intricate processes.
This improved interaction minimises errors, enhances the user’s understanding of the process, and helps them make more informed and appropriate trust decisions regarding security and privacy \cite{Lopez2020}. This helps organisations without extensive cybersecurity teams to conduct comprehensive assessments, identify vulnerabilities, and strengthen their security posture. Additionally, making CCMMs more accessible and understandable through user-centred design encourages a proactive attitude towards cybersecurity across the organisation \cite{Reynolds2021}. It demystifies the concept of cybersecurity maturity, making it a more tangible and achievable goal. This can lead to increased investment in cybersecurity initiatives, as decision-makers are better able to comprehend the value and necessity of advancing their organisation's cybersecurity maturity level.

\subsection{Addressing Implementation and Adoption Challenges}

In this section, we discuss strategies that can be employed to address the implementation and adoption challenges, offering a comprehensive roadmap for organisations to enhance their cybersecurity posture through the effective use of CCMMs.

\subsubsection{Simplifying CCMMs for better understanding and implementation.}  
Simplification and clarification of CCMMs can make them more accessible to organisations of different sizes and cybersecurity maturity levels \cite{Baikloy2020}. This involves creating clear, actionable steps and reducing complexity without compromising the comprehensiveness of the models. Simplified guidance documents, templates, and tools can assist organisations in understanding and implementing CCMMs more effectively.

\subsubsection{Integrating CCMMs with existing cybersecurity frameworks.}
To overcome integration challenges, CCMMs should be designed to complement and synergize with existing cybersecurity frameworks and standards (i.e. NIST Cybersecurity Framework, ISO 27001 and ISO 27002, Essential 8 (Australia), Cyber Essentials (UK)) \cite{Rea2017, Walter2014}. This requires a collaborative approach where developers of CCMMs work closely with the bodies governing other cybersecurity frameworks to create alignment and interoperability between systems \cite{Walter2014}. Such integration can provide organisations with a holistic cybersecurity approach that leverages the strengths of multiple frameworks.

\subsubsection{Enhancing the dynamic and actionable nature of CCMMs.}
CCMMs should be dynamic and continuously updated to reflect the latest cybersecurity threats and trends \cite{Le2016}. Incorporating elements of adaptability and actionability into the models can help organisations not just assess their current state but also implement practical improvements. This involves including clear, actionable steps for advancement at each maturity level and ensuring the models are flexible enough to adapt to new threats \cite{Gourisetti2020}.

\subsubsection{Building organisational engagement and responsibility.} 
Encouraging organisation-wide engagement in cybersecurity involves promoting a culture where cybersecurity is seen as a collective responsibility \cite{Buzdugan2022}. Leadership should actively advocate for cybersecurity as a critical component of the organisational strategy and champion the use of CCMMs as a tool for assessing and improving cybersecurity capabilities. To embed CCMMs into the organisational culture, it is essential to engage employees at all levels through targeted training programs that demystify the models and clarify their relevance to everyday operations. Engaging employees through training, gamification, and incentives can foster a proactive security culture across all departments, making the adoption and implementation of CCMMs a shared organisational goal \cite{AlMatari2021}.

\subsubsection{Resource allocation and skill development.}
Effective implementation of CCMMs requires a strategic approach to both resource allocation and skill development tailored specifically to advancing cybersecurity maturity. Organisations should prioritise the integration and use of CCMMs within their budgeting processes, ensuring that sufficient resources are allocated not only for initial adoption but also for ongoing assessments and improvements as outlined by the models \cite{Brown2018}. To address skill gaps that might hinder the effective use of CCMMs, organisations should invest in targeted training and development programs that are specifically designed to build the internal expertise necessary for understanding and applying the maturity models. Additionally, investing in these training and development programs can help build internal cybersecurity expertise, reducing reliance on external consultants.

\subsubsection{Regulatory adaptability and advocacy.}
To navigate the rapidly evolving regulatory landscape, organisations should adopt flexible cybersecurity frameworks that can quickly adapt to new regulations \cite{Adler2013, Brown2018}. Participating in policy discussions and advocacy can also help shape a regulatory environment that supports effective cybersecurity practices while being realistic about organisational capabilities \cite{Alayo2021}.

\section{Limitations}
\label{Section6-Limitations}

This SLR was conducted to provide a comprehensive overview of the topic by ensuring the reproducibility of the reported results in the literature. Nevertheless, while conducting the SLR, the returned articles from the search query are initially filtered based on their titles and abstracts using our inclusion and exclusion criteria, which could result in relevant articles being excluded during the selection phase. Therefore, to avoid such scenarios as much as possible, we performed an additional manual search using the backward snowballing approach, where we reviewed citations in the selected articles from the digital library search phase, as well as articles identified through searches in journals and conferences from top-tier venues \cite{Kitchenham04}. Following data collection, we conducted a thematic analysis to answer the research questions outlined in Section \ref{RQFormulation}. The first author performed the initial coding and theme identification. However, the generated codes and themes can be biased depending on the coder's experience, knowledge, and point of view. To reduce this bias, we considered the perspectives of all the authors when developing themes according to guidelines provided by Braun and Clerk \cite{Braun2012} as we have discussed in Section \ref{DataAnalysisSection}.

\vspace{-1em}

\section{Conclusion}
\label{Section7-Conclusion}

In the digital era, the pressing need for organisations to develop and maintain robust cybersecurity capabilities is undeniably critical to safeguard against cyberattacks \cite{Sebastiaan2017}. Cybersecurity Capability Maturity Models (CCMMs) have emerged as essential tools in guiding organisations through the complexities of establishing a secure posture \cite{Buzdugan2023}. These models offer structured pathways that systematically enable organisations to assess, plan, and enhance their cybersecurity measures - both technical and non-technical - including areas such as training, educational interventions, and organisational policies \cite{Curtis2014}. Yet, despite their invaluable frameworks, our SLR reveals significant limitations and challenges in the current CCMM landscape that necessitate a more evolved approach to meet the diverse and rapidly changing needs of the digital landscape.

The inherent limitations of existing CCMMs, such as their lack of flexibility and adaptability to evolving cyber threats, along with their inadequate alignment with the unique operational contexts of various organisations, highlight a critical gap in their conceptual (underlying principles and theory) and structural (how the domains, levels, and practices are articulated and interconnected to be actionable) design \cite{Aliyu2020, Akinsanya2019, Drivas2020, Le2017, Opeoluwa2019}. These models often provide guidance that is too generic or high-level to be directly actionable, leaving organisations, especially those with unique needs or those at different stages of cybersecurity maturity, at a disadvantage \cite{Adler2013}. Moreover, the challenges in implementing and adopting CCMMs, including resource constraints, a deficiency in cybersecurity awareness and culture, and the daunting complexity for accurate self-assessment, underscore the necessity for CCMMs to be more intuitive and tailored \cite{Walter2014, Brown2018}.

Addressing these gaps calls for the development of CCMMs that not only accommodate the dynamic nature of cyber threats but also offer tailored implementation paths. This involves the incorporation of domains and practices that provide a holistic perspective on cybersecurity maturity, effectively bridging the gap between theoretical models and practical, actionable strategies \cite{Hauck2011}. For organisations at varying levels of cybersecurity readiness, the introduction of tiered domains and practices is paramount. This differentiation ensures that each organisation can navigate the maturity model in a manner that aligns with its current capabilities and aspirations, making the journey towards cybersecurity excellence both achievable and measurable \cite{Spruit2014}.

The integration of meaningful metrics for continuous assessment and improvement is another crucial aspect of evolving CCMMs \cite{Karabacak2016}. By defining clear, quantifiable indicators of success, organisations can track their progress more effectively, identify areas for enhancement, and demonstrate the tangible benefits of their cybersecurity investments. This metrics-driven approach not only aids in the ongoing optimization of cybersecurity strategies but also fosters a culture of accountability and continuous improvement \cite{Akinsanya2019}.

Furthermore, the translation of CCMM elements into user-centred tooling support systems can address the critical challenge of complexity and accessibility. By designing tools and guidelines that are intuitive and user-centred, CCMMs become more accessible to a broader range of organisations, including those with limited cybersecurity expertise \cite{Lopez2020}. This user-centred design ensures that the process of assessing and enhancing cybersecurity maturity is not a daunting task but a manageable and engaging endeavour for all involved.

In conclusion, while CCMMs serve as foundational tools for enhancing organisational cybersecurity postures, our analysis underscores the imperative for these models to evolve. By embracing tailored implementation paths, introducing tiered domains and practices, integrating meaningful metrics, and developing user-centred tooling support systems, the next generation of CCMMs can overcome existing limitations and challenges. 

Future work should focus on the practical application of these evolved CCMMs in diverse organisational contexts, testing their effectiveness and adaptability in real-world settings. Additionally, investigating the integration of artificial intelligence and machine learning into CCMMs could provide organisations with predictive capabilities, enabling them to proactively address emerging cybersecurity challenges. Finally, future studies could examine the role of cross-sector collaboration in refining CCMMs, drawing on insights from various industries to create more universally applicable models. This evolution will enable CCMMs to provide more effective, dynamic, and inclusive frameworks that empower organisations to navigate the complexities of cybersecurity with confidence, ensuring their resilience against an ever-changing cyber threat landscape.

\bibliographystyle{ACM-Reference-Format}
\bibliography{sample-base}


\begin{thebibliography}{110}


\ifx \showCODEN    \undefined \def \showCODEN     #1{\unskip}     \fi
\ifx \showDOI      \undefined \def \showDOI       #1{#1}\fi
\ifx \showISBNx    \undefined \def \showISBNx     #1{\unskip}     \fi
\ifx \showISBNxiii \undefined \def \showISBNxiii  #1{\unskip}     \fi
\ifx \showISSN     \undefined \def \showISSN      #1{\unskip}     \fi
\ifx \showLCCN     \undefined \def \showLCCN      #1{\unskip}     \fi
\ifx \shownote     \undefined \def \shownote      #1{#1}          \fi
\ifx \showarticletitle \undefined \def \showarticletitle #1{#1}   \fi
\ifx \showURL      \undefined \def \showURL       {\relax}        \fi
\providecommand\bibfield[2]{#2}
\providecommand\bibinfo[2]{#2}
\providecommand\natexlab[1]{#1}
\providecommand\showeprint[2][]{arXiv:#2}

\bibitem[~({[n.\,d.]})]%
        {Hiscox2023}
\bibfield{author}{\bibinfo{person}{Hiscox  }.}
  \bibinfo{year}{[n.\,d.]}\natexlab{}.
\newblock \bibinfo{title}{Hiscox Cyber Readiness Report 2023}.
\newblock , \bibinfo{numpages}{25}~pages.
\newblock
\urldef\tempurl%
\url{https://www.hiscoxgroup.com/sites/group/files/documents/2023-10/Hiscox-Cyber-Readiness-Report-2023.pdf}
\showURL{%
\tempurl}


\bibitem[~(2021)]%
        {CybermaturityISACA}
\bibfield{author}{\bibinfo{person}{ISACA~Now  }.}
  \bibinfo{year}{2021}\natexlab{}.
\newblock \bibinfo{title}{Cybermaturity: A Path to Right-Sizing Cyber
  Practices}.
\newblock
\newblock
\urldef\tempurl%
\url{https://www.isaca.org/resources/news-and-trends/isaca-now-blog/2021/cybermaturity-a-path-to-right-sizing-cyber-practices}
\showURL{%
\tempurl}


\bibitem[Abdullahi~Garba et~al\mbox{.}(2020)]%
        {Abdullahi2020}
\bibfield{author}{\bibinfo{person}{Adamu Abdullahi~Garba},
  \bibinfo{person}{Aliyu Musa~Bade}, \bibinfo{person}{Muktar Yahuza}, {and}
  \bibinfo{person}{Ya’u Nuhu}.} \bibinfo{year}{2020}\natexlab{}.
\newblock \showarticletitle{Cybersecurity capability maturity models review and
  application domain}.
\newblock \bibinfo{journal}{\emph{International Journal of Engineering \&
  Technology}} \bibinfo{volume}{9}, \bibinfo{number}{3} (\bibinfo{year}{2020}),
  \bibinfo{pages}{779}.
\newblock
\showISSN{2227-524X}
\urldef\tempurl%
\url{https://doi.org/10.14419/ijet.v9i3.30719}
\showDOI{\tempurl}


\bibitem[Adler(2013)]%
        {Adler2013}
\bibfield{author}{\bibinfo{person}{Richard~M. Adler}.}
  \bibinfo{year}{2013}\natexlab{}.
\newblock \showarticletitle{A dynamic capability maturity model for improving
  cyber security}. In \bibinfo{booktitle}{\emph{2013 {IEEE} International
  Conference on Technologies for Homeland Security ({HST})}} (Waltham, {MA},
  {USA}, 2013-11). \bibinfo{publisher}{{IEEE}}, \bibinfo{address}{Waltham, MA,
  USA}, \bibinfo{pages}{230--235}.
\newblock
\showISBNx{978-1-4799-1535-4 978-1-4799-3963-3}
\urldef\tempurl%
\url{https://doi.org/10.1109/THS.2013.6699005}
\showDOI{\tempurl}


\bibitem[Agrafiotis et~al\mbox{.}(2018)]%
        {Agrafiotis2018}
\bibfield{author}{\bibinfo{person}{Ioannis Agrafiotis},
  \bibinfo{person}{Jason~RC Nurse}, \bibinfo{person}{Michael Goldsmith},
  \bibinfo{person}{Sadie Creese}, {and} \bibinfo{person}{David Upton}.}
  \bibinfo{year}{2018}\natexlab{}.
\newblock \showarticletitle{A taxonomy of cyber-harms: Defining the impacts of
  cyber-attacks and understanding how they propagate}.
\newblock \bibinfo{journal}{\emph{Journal of Cybersecurity}}
  \bibinfo{volume}{4}, \bibinfo{number}{1} (\bibinfo{year}{2018}),
  \bibinfo{pages}{tyy006}.
\newblock


\bibitem[Akinsanya et~al\mbox{.}(2019a)]%
        {Opeoluwa2019}
\bibfield{author}{\bibinfo{person}{Opeoluwa~Ore Akinsanya},
  \bibinfo{person}{Maria Papadaki}, {and} \bibinfo{person}{Lingfen Sun}.}
  \bibinfo{year}{2019}\natexlab{a}.
\newblock \showarticletitle{Current Cybersecurity Maturity Models: How
  Effective in Healthcare Cloud?}. In \bibinfo{booktitle}{\emph{Collaborative
  European Research Conference}}.
\newblock
\urldef\tempurl%
\url{https://api.semanticscholar.org/CorpusID:140242746}
\showURL{%
\tempurl}


\bibitem[Akinsanya et~al\mbox{.}(2019b)]%
        {Akinsanya2019}
\bibfield{author}{\bibinfo{person}{Opeoluwa~Ore Akinsanya},
  \bibinfo{person}{Maria Papadaki}, {and} \bibinfo{person}{Lingfen Sun}.}
  \bibinfo{year}{2019}\natexlab{b}.
\newblock \showarticletitle{Towards a maturity model for health-care cloud
  security (M $^{\textrm{2}}$ {HCS})}.
\newblock \bibinfo{journal}{\emph{Information \& Computer Security}}
  \bibinfo{volume}{28}, \bibinfo{number}{3} (\bibinfo{year}{2019}),
  \bibinfo{pages}{321--345}.
\newblock
\showISSN{2056-4961, 2056-4961}
\urldef\tempurl%
\url{https://doi.org/10.1108/ICS-05-2019-0060}
\showDOI{\tempurl}


\bibitem[Al-Matari et~al\mbox{.}(2021)]%
        {AlMatari2021}
\bibfield{author}{\bibinfo{person}{Osamah~M.M. Al-Matari},
  \bibinfo{person}{Iman~M.A. Helal}, \bibinfo{person}{Sherif~A. Mazen}, {and}
  \bibinfo{person}{Sherif Elhennawy}.} \bibinfo{year}{2021}\natexlab{}.
\newblock \showarticletitle{Adopting security maturity model to the
  organizations’ capability model}.
\newblock \bibinfo{journal}{\emph{Egyptian Informatics Journal}}
  \bibinfo{volume}{22}, \bibinfo{number}{2} (\bibinfo{year}{2021}),
  \bibinfo{pages}{193--199}.
\newblock
\showISSN{11108665}
\urldef\tempurl%
\url{https://doi.org/10.1016/j.eij.2020.08.001}
\showDOI{\tempurl}


\bibitem[Alayo et~al\mbox{.}(2021)]%
        {Alayo2021}
\bibfield{author}{\bibinfo{person}{Jorge~Gutierrez Alayo},
  \bibinfo{person}{Paul~Necochea Mendoza}, \bibinfo{person}{Jimmy
  Armas-Aguirre}, {and} \bibinfo{person}{Juan~Madrid Molina}.}
  \bibinfo{year}{2021}\natexlab{}.
\newblock \showarticletitle{Cybersecurity maturity model for providing services
  in the financial sector in Peru}. In \bibinfo{booktitle}{\emph{2021 Congreso
  Internacional de Innovaci{\'o}n y Tendencias en Ingenier{\'\i}a (CONIITI)}}.
  IEEE, \bibinfo{pages}{1--4}.
\newblock


\bibitem[Aliyu et~al\mbox{.}(2020)]%
        {Aliyu2020}
\bibfield{author}{\bibinfo{person}{Aliyu Aliyu}, \bibinfo{person}{Leandros
  Maglaras}, \bibinfo{person}{Ying He}, \bibinfo{person}{Iryna Yevseyeva},
  \bibinfo{person}{Eerke Boiten}, \bibinfo{person}{Allan Cook}, {and}
  \bibinfo{person}{Helge Janicke}.} \bibinfo{year}{2020}\natexlab{}.
\newblock \showarticletitle{A holistic cybersecurity maturity assessment
  framework for higher education institutions in the United Kingdom}.
\newblock \bibinfo{journal}{\emph{Applied Sciences}} \bibinfo{volume}{10},
  \bibinfo{number}{10} (\bibinfo{year}{2020}), \bibinfo{pages}{3660}.
\newblock


\bibitem[Aljundi et~al\mbox{.}(2023)]%
        {Aljundi2023}
\bibfield{author}{\bibinfo{person}{Ibrahem Aljundi}, \bibinfo{person}{Morad
  Rawashdeh}, \bibinfo{person}{Mustafa Al-Fayoumi}, \bibinfo{person}{Amer
  Al-Badarneh}, {and} \bibinfo{person}{Qasem~Abu Al-Haija}.}
  \bibinfo{year}{2023}\natexlab{}.
\newblock \showarticletitle{Protecting Critical National Infrastructures: An
  Overview of Cyberattacks and Countermeasures}. In
  \bibinfo{booktitle}{\emph{International conference on WorldS4}}. Springer,
  \bibinfo{pages}{295--317}.
\newblock


\bibitem[Almuhammadi and Alsaleh(2017)]%
        {Almuhammadi2017}
\bibfield{author}{\bibinfo{person}{Sultan Almuhammadi} {and}
  \bibinfo{person}{Majeed Alsaleh}.} \bibinfo{year}{2017}\natexlab{}.
\newblock \showarticletitle{Information security maturity model for NIST cyber
  security framework}.
\newblock \bibinfo{journal}{\emph{Computer Science \& Information Technology
  (CS \& IT)}} \bibinfo{volume}{7}, \bibinfo{number}{3} (\bibinfo{year}{2017}),
  \bibinfo{pages}{51--62}.
\newblock


\bibitem[Alqudhaibi et~al\mbox{.}(2023)]%
        {Alqudhaibi2023}
\bibfield{author}{\bibinfo{person}{Adel Alqudhaibi}, \bibinfo{person}{Majed
  Albarrak}, \bibinfo{person}{Abdulmohsan Aloseel}, \bibinfo{person}{Sandeep
  Jagtap}, {and} \bibinfo{person}{Konstantinos Salonitis}.}
  \bibinfo{year}{2023}\natexlab{}.
\newblock \showarticletitle{Predicting cybersecurity threats in critical
  infrastructure for industry 4.0: a proactive approach based on attacker
  motivations}.
\newblock \bibinfo{journal}{\emph{Sensors}} \bibinfo{volume}{23},
  \bibinfo{number}{9} (\bibinfo{year}{2023}), \bibinfo{pages}{4539}.
\newblock


\bibitem[Altaha and Rahman(2023)]%
        {Altaha2023}
\bibfield{author}{\bibinfo{person}{Safa Altaha} {and}
  \bibinfo{person}{MM~Hafizur Rahman}.} \bibinfo{year}{2023}\natexlab{}.
\newblock \showarticletitle{A mini literature review on integrating
  cybersecurity for business continuity}. In \bibinfo{booktitle}{\emph{2023
  International Conference on Artificial Intelligence in Information and
  Communication (ICAIIC)}}. IEEE, \bibinfo{pages}{353--359}.
\newblock


\bibitem[Antoniazzi and Salihudin(2023)]%
        {Antoniazzi2023}
\bibfield{author}{\bibinfo{person}{Giovanna Antoniazzi} {and}
  \bibinfo{person}{Saiful Salihudin}.} \bibinfo{year}{2023}\natexlab{}.
\newblock \bibinfo{title}{State of the Connected World 2023 Edition}.
\newblock , \bibinfo{numpages}{49}~pages.
\newblock
\urldef\tempurl%
\url{https://www3.weforum.org/docs/WEF_State_of_the_Connected_World_2023_Edition.pdf}
\showURL{%
\tempurl}


\bibitem[Awa et~al\mbox{.}(2017)]%
        {Awa2017}
\bibfield{author}{\bibinfo{person}{Hart~O Awa}, \bibinfo{person}{Ojiabo~Ukoha
  Ojiabo}, {and} \bibinfo{person}{Longlife~E Orokor}.}
  \bibinfo{year}{2017}\natexlab{}.
\newblock \showarticletitle{Integrated technology-organization-environment
  (TOE) taxonomies for technology adoption}.
\newblock \bibinfo{journal}{\emph{Journal of Enterprise Information
  Management}} \bibinfo{volume}{30}, \bibinfo{number}{6}
  (\bibinfo{year}{2017}), \bibinfo{pages}{893--921}.
\newblock


\bibitem[Bahuguna et~al\mbox{.}(2019)]%
        {Bahuguna2019}
\bibfield{author}{\bibinfo{person}{Ashutosh Bahuguna},
  \bibinfo{person}{Raj~Kishor Bisht}, {and} \bibinfo{person}{Jeetendra Pande}.}
  \bibinfo{year}{2019}\natexlab{}.
\newblock \showarticletitle{Assessing cybersecurity maturity of organizations:
  An empirical investigation in the Indian context}.
\newblock \bibinfo{journal}{\emph{Information Security Journal: A Global
  Perspective}} \bibinfo{volume}{28}, \bibinfo{number}{6}
  (\bibinfo{year}{2019}), \bibinfo{pages}{164--177}.
\newblock


\bibitem[Baikloy et~al\mbox{.}(2020)]%
        {Baikloy2020}
\bibfield{author}{\bibinfo{person}{Ekkachat Baikloy}, \bibinfo{person}{Prasong
  Praneetpolgrang}, {and} \bibinfo{person}{Nivet Jirawichitchai}.}
  \bibinfo{year}{2020}\natexlab{}.
\newblock \showarticletitle{Development of Cyber Resilient Capability Maturity
  Model for Cloud Computing Services.}
\newblock \bibinfo{journal}{\emph{TEM Journal}} \bibinfo{volume}{9},
  \bibinfo{number}{3} (\bibinfo{year}{2020}).
\newblock


\bibitem[Barclay(2014)]%
        {Barclay2014}
\bibfield{author}{\bibinfo{person}{Corlane Barclay}.}
  \bibinfo{year}{2014}\natexlab{}.
\newblock \showarticletitle{Sustainable security advantage in a changing
  environment: The Cybersecurity Capability Maturity Model
  ({CM}$^{\textrm{2}}$)}. In \bibinfo{booktitle}{\emph{Proceedings of the 2014
  {ITU} kaleidoscope academic conference: Living in a converged world -
  Impossible without standards?}} (Saint-Petersburg, Russia, 2014-06).
  \bibinfo{publisher}{{IEEE}}, \bibinfo{pages}{275--282}.
\newblock
\showISBNx{978-92-61-14421-0}
\urldef\tempurl%
\url{https://doi.org/10.1109/Kaleidoscope.2014.6858466}
\showDOI{\tempurl}


\bibitem[Barnes and Daim(2022)]%
        {Barnes2022}
\bibfield{author}{\bibinfo{person}{Bridget Barnes} {and}
  \bibinfo{person}{Tugrul Daim}.} \bibinfo{year}{2022}\natexlab{}.
\newblock \showarticletitle{Information security maturity model for healthcare
  organizations in the United States}.
\newblock \bibinfo{journal}{\emph{IEEE Transactions on Engineering Management}}
  (\bibinfo{year}{2022}).
\newblock


\bibitem[Bellasio et~al\mbox{.}(2018)]%
        {Bellasio2018}
\bibfield{author}{\bibinfo{person}{Jacopo Bellasio}, \bibinfo{person}{Richard
  Flint}, \bibinfo{person}{Nathan Ryan}, \bibinfo{person}{S Sondergaard},
  \bibinfo{person}{Cristina~Gonzalez Monsalve}, \bibinfo{person}{Arya~Sofia
  Meranto}, {and} \bibinfo{person}{Anna Knack}.}
  \bibinfo{year}{2018}\natexlab{}.
\newblock \showarticletitle{Developing Cybersecurity Capacity}.
\newblock \bibinfo{journal}{\emph{RAND Europe corporation}}
  (\bibinfo{year}{2018}).
\newblock


\bibitem[Bhattacharya et~al\mbox{.}(2022)]%
        {Bhattacharya2022}
\bibfield{author}{\bibinfo{person}{Souradeep Bhattacharya},
  \bibinfo{person}{Burhan Hyder}, {and} \bibinfo{person}{Manimaran
  Govindarasu}.} \bibinfo{year}{2022}\natexlab{}.
\newblock \showarticletitle{{ICS}-{CTM}2: Industrial Control System
  Cybersecurity Testbed Maturity Model}. In \bibinfo{booktitle}{\emph{2022
  Resilience Week ({RWS})}} (National Harbor, {MD}, {USA}, 2022-09-26).
  \bibinfo{publisher}{{IEEE}}, \bibinfo{pages}{1--6}.
\newblock
\showISBNx{978-1-66548-819-8}
\urldef\tempurl%
\url{https://doi.org/10.1109/RWS55399.2022.9984023}
\showDOI{\tempurl}


\bibitem[Booth et~al\mbox{.}(2012)]%
        {Booth2012}
\bibfield{author}{\bibinfo{person}{Andrew Booth}, \bibinfo{person}{Anthea
  Sutton}, \bibinfo{person}{Mark Clowes}, {and} \bibinfo{person}{Marrissa
  Martyn-St~James}.} \bibinfo{year}{2012}\natexlab{}.
\newblock \bibinfo{title}{Systematic Approaches to a Successful Literature
  Review}.
\newblock
\newblock
\urldef\tempurl%
\url{https://uk.sagepub.com/en-gb/eur/systematic-approaches-to-a-successful-literature-review/book270933#description}
\showURL{%
\tempurl}


\bibitem[Braun and Clarke(2012)]%
        {Braun2012}
\bibfield{author}{\bibinfo{person}{Virginia Braun} {and}
  \bibinfo{person}{Victoria Clarke}.} \bibinfo{year}{2012}\natexlab{}.
\newblock \bibinfo{booktitle}{\emph{Thematic analysis.}}
\newblock \bibinfo{publisher}{American Psychological Association}.
\newblock


\bibitem[Brooks(2024)]%
        {Brooks2024}
\bibfield{author}{\bibinfo{person}{Khristopher~J Brooks}.}
  \bibinfo{year}{2024}\natexlab{}.
\newblock \bibinfo{title}{UnitedHealth says Change Healthcare cyberattack cost
  it \$872 million}.
\newblock
\newblock
\urldef\tempurl%
\url{https://www.cbsnews.com/news/unitedhealth-cyberattack-change-healthcare-hack-ransomware/}
\showURL{%
\tempurl}


\bibitem[Brown(2018)]%
        {Brown2018}
\bibfield{author}{\bibinfo{person}{Rafael~Dean Brown}.}
  \bibinfo{year}{2018}\natexlab{}.
\newblock \showarticletitle{Towards a Qatar Cybersecurity Capability Maturity
  Model with a Legislative Framework}.
\newblock \bibinfo{journal}{\emph{International Review of Law}}
  (\bibinfo{year}{2018}).
\newblock


\bibitem[Buzdugan and Capatana(2022)]%
        {Buzdugan2022}
\bibfield{author}{\bibinfo{person}{Aurelian Buzdugan} {and}
  \bibinfo{person}{Gheorghe Capatana}.} \bibinfo{year}{2022}\natexlab{}.
\newblock \showarticletitle{Cyber security maturity model for critical
  infrastructures}. In \bibinfo{booktitle}{\emph{Education, Research and
  Business Technologies: Proceedings of 20th International Conference on
  Informatics in Economy (IE 2021)}}. Springer, \bibinfo{pages}{225--236}.
\newblock


\bibitem[Buzdugan and C{\u{a}}p{\u{a}}ț{\^a}n{\u{a}}(2023)]%
        {Buzdugan2023}
\bibfield{author}{\bibinfo{person}{Aurelian Buzdugan} {and}
  \bibinfo{person}{Gheorghe C{\u{a}}p{\u{a}}ț{\^a}n{\u{a}}}.}
  \bibinfo{year}{2023}\natexlab{}.
\newblock \showarticletitle{The Trends in Cybersecurity Maturity Models}. In
  \bibinfo{booktitle}{\emph{Education, Research and Business Technologies:
  Proceedings of 21st International Conference on Informatics in Economy (IE
  2022)}}. Springer, \bibinfo{pages}{217--228}.
\newblock


\bibitem[Christopher(2014)]%
        {ESC2M2}
\bibfield{author}{\bibinfo{person}{Jason~D. Christopher}.}
  \bibinfo{year}{2014}\natexlab{}.
\newblock \bibinfo{booktitle}{\emph{Electricity Subsector Cybersecurity
  Capability Maturity Model (ES-C2M2)}}.
\newblock \bibinfo{type}{{T}echnical {R}eport}. \bibinfo{institution}{U.S.
  Department of Energy}. \bibinfo{pages}{1--89} pages.
\newblock


\bibitem[Christopher et~al\mbox{.}(2014a)]%
        {C2M2}
\bibfield{author}{\bibinfo{person}{Jason~D. Christopher}, \bibinfo{person}{Dale
  Gonzalez}, \bibinfo{person}{David~W. White}, \bibinfo{person}{James Stevens},
  \bibinfo{person}{Julie Grundman}, \bibinfo{person}{Nader Mehravari},
  \bibinfo{person}{Pamela Curtis}, {and} \bibinfo{person}{Tom Dolan}.}
  \bibinfo{year}{2014}\natexlab{a}.
\newblock \bibinfo{booktitle}{\emph{Cybersecurity Capability Maturity Model
  (C2M2)}}.
\newblock \bibinfo{type}{{T}echnical {R}eport}. \bibinfo{institution}{U.S.
  Department of Energy}. \bibinfo{pages}{1--76} pages.
\newblock


\bibitem[Christopher et~al\mbox{.}(2014b)]%
        {ONGC2M2}
\bibfield{author}{\bibinfo{person}{Jason~D. Christopher}, \bibinfo{person}{Beth
  Lemke}, \bibinfo{person}{Dan Strachan}, \bibinfo{person}{David~W. White},
  \bibinfo{person}{Drew Kittey}, \bibinfo{person}{Dustin Brooks},
  \bibinfo{person}{Evon Sallee}, \bibinfo{person}{Jack~Eisenhauer Eisenhauer},
  \bibinfo{person}{Jack Whitsitt}, \bibinfo{person}{James~W. Sample},
  \bibinfo{person}{Jim Fisher}, \bibinfo{person}{John~S. Townsend},
  \bibinfo{person}{Jonathan Murphy}, \bibinfo{person}{Keith Dodrill},
  \bibinfo{person}{Keith~H. Herndon}, \bibinfo{person}{Kelley Bray},
  \bibinfo{person}{Kimberly Denbow}, \bibinfo{person}{Lindsay Kishter},
  \bibinfo{person}{Lisa Kaiser}, \bibinfo{person}{Matthew Harper},
  \bibinfo{person}{Paul Skare}, \bibinfo{person}{Penny Wolter},
  \bibinfo{person}{Peter Sindt}, \bibinfo{person}{R.~Peter Weaver},
  \bibinfo{person}{Robert Mims}, \bibinfo{person}{Robert Mims},
  \bibinfo{person}{Scott~M. Baron}, \bibinfo{person}{Scott vonFischer},
  \bibinfo{person}{Scott Womer}, \bibinfo{person}{Seamus Stack},
  \bibinfo{person}{Suzanne Lemieux}, \bibinfo{person}{Tamara Lance},
  \bibinfo{person}{Terry Boss}, {and} \bibinfo{person}{Thomas Whitmore}.}
  \bibinfo{year}{2014}\natexlab{b}.
\newblock \bibinfo{booktitle}{\emph{Oil and Natural Gas Subsector Cybersecurity
  Capability Maturity Model (ONG-C2M2)}}.
\newblock \bibinfo{type}{{T}echnical {R}eport}. \bibinfo{institution}{U.S.
  Department of Energy}. \bibinfo{pages}{1--83} pages.
\newblock


\bibitem[Cruzes and Dyba(2011)]%
        {Cruzes2011}
\bibfield{author}{\bibinfo{person}{Daniela~S Cruzes} {and}
  \bibinfo{person}{Tore Dyba}.} \bibinfo{year}{2011}\natexlab{}.
\newblock \showarticletitle{Recommended steps for thematic synthesis in
  software engineering}. In \bibinfo{booktitle}{\emph{2011 international
  symposium on empirical software engineering and measurement}}. IEEE,
  \bibinfo{pages}{275--284}.
\newblock


\bibitem[Curtis et~al\mbox{.}(2014)]%
        {Curtis2014}
\bibfield{author}{\bibinfo{person}{Pamela Curtis}, \bibinfo{person}{Nader
  Mehravari}, {and} \bibinfo{person}{James Stevens}.}
  \bibinfo{year}{2014}\natexlab{}.
\newblock \bibinfo{booktitle}{\emph{Cybersecurity Capability Maturity Model for
  Information Technology Services (C2M2 for {IT} Services), Version 1.0}}.
\newblock \bibinfo{type}{Carnegie Mellon University ,Software Engineering
  Institute}. \bibinfo{institution}{CERT, Carnegie Mellon University},
  \bibinfo{address}{Carnegie Mellon University,Software Engineering Institute}.
\newblock


\bibitem[Curtis and Mehravari(2015)]%
        {Curtis2015}
\bibfield{author}{\bibinfo{person}{Pamela~D. Curtis} {and}
  \bibinfo{person}{Nader Mehravari}.} \bibinfo{year}{2015}\natexlab{}.
\newblock \showarticletitle{Evaluating and improving cybersecurity capabilities
  of the energy critical infrastructure}. In \bibinfo{booktitle}{\emph{2015
  {IEEE} International Symposium on Technologies for Homeland Security
  ({HST})}} (Waltham, {MA}, {USA}, 2015-04). \bibinfo{publisher}{{IEEE}},
  \bibinfo{pages}{1--6}.
\newblock
\showISBNx{978-1-4799-1737-2}
\urldef\tempurl%
\url{https://doi.org/10.1109/THS.2015.7225323}
\showDOI{\tempurl}


\bibitem[Daraghmeh and Brown(2021)]%
        {Daraghmeh2021}
\bibfield{author}{\bibinfo{person}{Rania Daraghmeh} {and}
  \bibinfo{person}{Raymond Brown}.} \bibinfo{year}{2021}\natexlab{}.
\newblock \showarticletitle{A Big Data Maturity Model for Electronic Health
  Records in Hospitals}. In \bibinfo{booktitle}{\emph{2021 International
  Conference on Information Technology ({ICIT})}} (Amman, Jordan, 2021-07-14).
  \bibinfo{publisher}{{IEEE}}, \bibinfo{pages}{826--833}.
\newblock
\showISBNx{978-1-66542-870-5}
\urldef\tempurl%
\url{https://doi.org/10.1109/ICIT52682.2021.9491781}
\showDOI{\tempurl}


\bibitem[Divadari et~al\mbox{.}(2023)]%
        {Divadari2023}
\bibfield{author}{\bibinfo{person}{Satyavathi Divadari}, \bibinfo{person}{J
  Surya~Prasad}, {and} \bibinfo{person}{Prasad Honnavalli}.}
  \bibinfo{year}{2023}\natexlab{}.
\newblock \showarticletitle{Managing data protection and privacy on cloud}. In
  \bibinfo{booktitle}{\emph{Proceedings of 3rd International Conference on
  Recent Trends in Machine Learning, IoT, Smart Cities and Applications: ICMISC
  2022}}. Springer, \bibinfo{pages}{383--396}.
\newblock


\bibitem[Drivas et~al\mbox{.}(2020)]%
        {Drivas2020}
\bibfield{author}{\bibinfo{person}{George Drivas}, \bibinfo{person}{Argyro
  Chatzopoulou}, \bibinfo{person}{Leandros Maglaras}, \bibinfo{person}{Costas
  Lambrinoudakis}, \bibinfo{person}{Allan Cook}, {and} \bibinfo{person}{Helge
  Janicke}.} \bibinfo{year}{2020}\natexlab{}.
\newblock \showarticletitle{A NIS Directive Compliant Cybersecurity Maturity
  Assessment Framework}. In \bibinfo{booktitle}{\emph{2020 IEEE 44th Annual
  Computers, Software, and Applications Conference (COMPSAC)}}. IEEE,
  \bibinfo{pages}{1641--1646}.
\newblock


\bibitem[Dube and Mohanty(2020)]%
        {Dube2020}
\bibfield{author}{\bibinfo{person}{Durga~Prasad Dube} {and}
  \bibinfo{person}{R.P. Mohanty}.} \bibinfo{year}{2020}\natexlab{}.
\newblock \showarticletitle{Towards development of a cyber security capability
  maturity model}.
\newblock \bibinfo{journal}{\emph{International Journal of Business Information
  Systems}} \bibinfo{volume}{34}, \bibinfo{number}{1} (\bibinfo{year}{2020}),
  \bibinfo{pages}{104}.
\newblock
\showISSN{1746-0972, 1746-0980}
\urldef\tempurl%
\url{https://doi.org/10.1504/IJBIS.2020.106800}
\showDOI{\tempurl}


\bibitem[Dube and Mohanty(2021)]%
        {Dube2021}
\bibfield{author}{\bibinfo{person}{Durga~Prasad Dube} {and} \bibinfo{person}{RP
  Mohanty}.} \bibinfo{year}{2021}\natexlab{}.
\newblock \showarticletitle{The application of cyber security capability
  maturity model to identify the impact of internal efficiency factors on the
  external effectiveness of cyber security}.
\newblock \bibinfo{journal}{\emph{International Journal of Business Information
  Systems}} \bibinfo{volume}{38}, \bibinfo{number}{3} (\bibinfo{year}{2021}),
  \bibinfo{pages}{367--392}.
\newblock


\bibitem[Ebrahimnezhad and Sepehri(2023)]%
        {Ebrahimnezhad2023}
\bibfield{author}{\bibinfo{person}{Mohammad Ebrahimnezhad} {and}
  \bibinfo{person}{Mehran Sepehri}.} \bibinfo{year}{2023}\natexlab{}.
\newblock \showarticletitle{Cyber Resiliency in Electric Power Industry Based
  on the Maturity Model}. In \bibinfo{booktitle}{\emph{Cybersecurity in the Age
  of Smart Societies: Proceedings of the 14th International Conference on
  Global Security, Safety and Sustainability, London, September 2022}}.
  Springer, \bibinfo{pages}{411--420}.
\newblock


\bibitem[European Conference on Cyber Warfare and Security (ECCWS)(2017)]%
        {Jacobs2017}
European Conference on Cyber Warfare and Security (ECCWS)
  \bibinfo{year}{2017}\natexlab{}.
\newblock \bibinfo{booktitle}{\emph{Towards a National Cybersecurity Capability
  Development Model}}. European Conference on Cyber Warfare and Security
  (ECCWS).
\newblock
\urldef\tempurl%
\url{https://researchspace.csir.co.za/dspace/handle/10204/9458}
\showURL{%
\tempurl}


\bibitem[Fink(2019)]%
        {Fink2019}
\bibfield{author}{\bibinfo{person}{Arlene Fink}.}
  \bibinfo{year}{2019}\natexlab{}.
\newblock \bibinfo{booktitle}{\emph{Conducting research literature reviews:
  From the internet to paper}}.
\newblock \bibinfo{publisher}{Sage publications}.
\newblock


\bibitem[Gcaza and Von~Solms(2017)]%
        {Gcaza2017}
\bibfield{author}{\bibinfo{person}{Noluxolo Gcaza} {and}
  \bibinfo{person}{Rossouw Von~Solms}.} \bibinfo{year}{2017}\natexlab{}.
\newblock \showarticletitle{Cybersecurity culture: An ill-defined problem}. In
  \bibinfo{booktitle}{\emph{Information Security Education for a Global Digital
  Society: 10th IFIP WG 11.8 World Conference, WISE 10, Rome, Italy, May 29-31,
  2017, Proceedings 10}}. Springer, \bibinfo{pages}{98--109}.
\newblock


\bibitem[Georgiadou et~al\mbox{.}(2022)]%
        {Georgiadou2022}
\bibfield{author}{\bibinfo{person}{Anna Georgiadou}, \bibinfo{person}{Spiros
  Mouzakitis}, \bibinfo{person}{Kanaris Bounas}, {and}
  \bibinfo{person}{Dimitrios Askounis}.} \bibinfo{year}{2022}\natexlab{}.
\newblock \showarticletitle{A cyber-security culture framework for assessing
  organization readiness}.
\newblock \bibinfo{journal}{\emph{Journal of Computer Information Systems}}
  \bibinfo{volume}{62}, \bibinfo{number}{3} (\bibinfo{year}{2022}),
  \bibinfo{pages}{452--462}.
\newblock


\bibitem[Georgiev(2021)]%
        {Georgiev2021}
\bibfield{author}{\bibinfo{person}{Venelin Georgiev}.}
  \bibinfo{year}{2021}\natexlab{}.
\newblock \showarticletitle{Comparative Analysis of the Cyber Security
  Capabilities Maturity Models}.
\newblock \bibinfo{journal}{\emph{Godishnik na UNSS}}  \bibinfo{volume}{2}
  (\bibinfo{year}{2021}), \bibinfo{pages}{31--42}.
\newblock
\urldef\tempurl%
\url{https://EconPapers.repec.org/RePEc:nwe:godish:y:2021:i:2:p:31-42}
\showURL{%
\tempurl}


\bibitem[Ghaffari and Arabsorkhi(2018)]%
        {Ghaffari2018}
\bibfield{author}{\bibinfo{person}{Fariba Ghaffari} {and}
  \bibinfo{person}{Abouzar Arabsorkhi}.} \bibinfo{year}{2018}\natexlab{}.
\newblock \showarticletitle{A New Adaptive Cyber Security Capability Maturity
  Model}. In \bibinfo{booktitle}{\emph{2018 9th International Symposium on
  Telecommunications (IST)}}. IEEE, \bibinfo{pages}{298--304}.
\newblock


\bibitem[Glantz et~al\mbox{.}(2016)]%
        {Glantz2016}
\bibfield{author}{\bibinfo{person}{Clifford Glantz}, \bibinfo{person}{Sriram
  Somasundaram}, \bibinfo{person}{Michael Mylrea}, \bibinfo{person}{Ron
  Underhill}, {and} \bibinfo{person}{Andrew Nicholls}.}
  \bibinfo{year}{2016}\natexlab{}.
\newblock \showarticletitle{Evaluating the maturity of cybersecurity programs
  for building control systems}.
\newblock \bibinfo{journal}{\emph{US Department of Energy Office of Scientific
  and Technical Information}} (\bibinfo{year}{2016}).
\newblock


\bibitem[Gourisetti et~al\mbox{.}(2019)]%
        {Gourisetti2019}
\bibfield{author}{\bibinfo{person}{Sri Nikhil~Gupta Gourisetti},
  \bibinfo{person}{Scott Mix}, \bibinfo{person}{Michael Mylrea},
  \bibinfo{person}{Christopher Bonebrake}, {and} \bibinfo{person}{Md
  Touhiduzzaman}.} \bibinfo{year}{2019}\natexlab{}.
\newblock \showarticletitle{Secure Design and Development Cybersecurity
  Capability Maturity Model ({SD}2-C2M2): Next-Generation Cyber Resilience by
  Design}. In \bibinfo{booktitle}{\emph{Proceedings of the Northwest
  Cybersecurity Symposium}} (Richland {WA} {USA}, 2019-04-08).
  \bibinfo{publisher}{{ACM}}, \bibinfo{pages}{1--9}.
\newblock
\showISBNx{978-1-4503-6614-4}
\urldef\tempurl%
\url{https://doi.org/10.1145/3332448.3332461}
\showDOI{\tempurl}


\bibitem[Gourisetti et~al\mbox{.}(2020)]%
        {Gourisetti2020}
\bibfield{author}{\bibinfo{person}{Sri Nikhil~Gupta Gourisetti},
  \bibinfo{person}{Michael Mylrea}, {and} \bibinfo{person}{Hirak Patangia}.}
  \bibinfo{year}{2020}\natexlab{}.
\newblock \showarticletitle{Cybersecurity vulnerability mitigation framework
  through empirical paradigm: Enhanced prioritized gap analysis}.
\newblock \bibinfo{journal}{\emph{Future Generation Computer Systems}}
  \bibinfo{volume}{105} (\bibinfo{year}{2020}), \bibinfo{pages}{410--431}.
\newblock
\showISSN{0167739X}
\urldef\tempurl%
\url{https://doi.org/10.1016/j.future.2019.12.018}
\showDOI{\tempurl}


\bibitem[Government(2022)]%
        {CSStrategy}
\bibfield{author}{\bibinfo{person}{HM Government}.}
  \bibinfo{year}{2022}\natexlab{}.
\newblock \showarticletitle{Government Cyber Secuirty Strategy 2022-2030}.
\newblock \bibinfo{journal}{\emph{HM Government}} \bibinfo{volume}{1},
  \bibinfo{number}{1} (\bibinfo{year}{2022}), \bibinfo{pages}{1--84}.
\newblock
\urldef\tempurl%
\url{https://assets.publishing.service.gov.uk/government/uploads/system/uploads/attachment_data/file/1049825/government-cyber-security-strategy.pdf}
\showURL{%
\tempurl}


\bibitem[Hauck et~al\mbox{.}(2011)]%
        {Hauck2011}
\bibfield{author}{\bibinfo{person}{Jean Carlo~Rossa Hauck},
  \bibinfo{person}{Christiane~Gresse Von~Wangenheim}, \bibinfo{person}{Fergal
  Mc~Caffery}, {and} \bibinfo{person}{Luigi Buglione}.}
  \bibinfo{year}{2011}\natexlab{}.
\newblock \showarticletitle{Proposing an ISO/IEC 15504-2 compliant method for
  process capability/maturity models customization}. In
  \bibinfo{booktitle}{\emph{Product-Focused Software Process Improvement: 12th
  International Conference, PROFES 2011, Torre Canne, Italy, June 20-22, 2011.
  Proceedings 12}}. Springer, \bibinfo{pages}{44--58}.
\newblock


\bibitem[Höpken et~al\mbox{.}(2021)]%
        {Wolfram2021}
\bibfield{author}{\bibinfo{person}{Wolfram Höpken}, \bibinfo{person}{Tobias
  Eberle}, \bibinfo{person}{Matthias Fuchs}, {and} \bibinfo{person}{Maria
  Lexhagen}.} \bibinfo{year}{2021}\natexlab{}.
\newblock \showarticletitle{Improving Tourist Arrival Prediction: A Big Data
  and Artificial Neural Network Approach}.
\newblock \bibinfo{journal}{\emph{Journal of Travel Research}}
  \bibinfo{volume}{60}, \bibinfo{number}{5} (\bibinfo{year}{2021}),
  \bibinfo{pages}{998--1017}.
\newblock
\urldef\tempurl%
\url{https://doi.org/10.1177/0047287520921244}
\showDOI{\tempurl}
\showeprint{https://doi.org/10.1177/0047287520921244}


\bibitem[Imran et~al\mbox{.}(2022)]%
        {Imran2022}
\bibfield{author}{\bibinfo{person}{Huma Imran}, \bibinfo{person}{Mohamed
  Salama}, \bibinfo{person}{Colin Turner}, {and} \bibinfo{person}{Sherif
  Fattah}.} \bibinfo{year}{2022}\natexlab{}.
\newblock \showarticletitle{Cybersecurity risk management frameworks in the oil
  and gas sector: A systematic literature review}. In
  \bibinfo{booktitle}{\emph{Future of Information and Communication
  Conference}}. Springer, \bibinfo{pages}{871--894}.
\newblock


\bibitem[Jaquire and von Solms(2017a)]%
        {Sebastiaan2017}
\bibfield{author}{\bibinfo{person}{Victor Jaquire} {and}
  \bibinfo{person}{Sebastiaan von Solms}.} \bibinfo{year}{2017}\natexlab{a}.
\newblock \showarticletitle{Cultivating a cyber counterintelligence maturity
  model}. In \bibinfo{booktitle}{\emph{ECCWS 2017 16th European Conference on
  Cyber Warfare and Security}}. Academic Conferences and Publishing Limited,
  \bibinfo{pages}{176}.
\newblock


\bibitem[Jaquire and von Solms(2017b)]%
        {Victor2017}
\bibfield{author}{\bibinfo{person}{Victor Jaquire} {and}
  \bibinfo{person}{Sebastian von Solms}.} \bibinfo{year}{2017}\natexlab{b}.
\newblock \showarticletitle{Developing a cyber counterintelligence maturity
  model for developing countries}. In \bibinfo{booktitle}{\emph{2017 IST-Africa
  Week Conference (IST-Africa)}}. IEEE, \bibinfo{pages}{1--8}.
\newblock


\bibitem[Jaquire and von Solms(2017c)]%
        {Jaquire2017}
\bibfield{author}{\bibinfo{person}{Victor Jaquire} {and}
  \bibinfo{person}{Sebastiaan von Solms}.} \bibinfo{year}{2017}\natexlab{c}.
\newblock \showarticletitle{Towards a cyber counterintelligence maturity
  model}. In \bibinfo{booktitle}{\emph{Proceedings of the 12th International
  Conference on Cyber Warfare and Security, AR Bryant \& RF Mills (eds), Wright
  State University, Air Force Institute of Technology, Dayton, OH, US}}.
  \bibinfo{pages}{432--40}.
\newblock


\bibitem[Jeong et~al\mbox{.}(2022)]%
        {Jeong2022}
\bibfield{author}{\bibinfo{person}{Jongkil Jeong}, \bibinfo{person}{Marthie
  Grobler}, {and} \bibinfo{person}{MAP Chamikara}.}
  \bibinfo{year}{2022}\natexlab{}.
\newblock \showarticletitle{Simplifying Cyber Security Maturity Models through
  National Culture: A Fuzzy Logic Approach.}. In
  \bibinfo{booktitle}{\emph{HICSS}}. \bibinfo{pages}{1--10}.
\newblock


\bibitem[Jones and Smyth(2004)]%
        {Jones2004}
\bibfield{author}{\bibinfo{person}{Leanne~V Jones} {and}
  \bibinfo{person}{Rosalind~L Smyth}.} \bibinfo{year}{2004}\natexlab{}.
\newblock \showarticletitle{How to perform a literature search}.
\newblock \bibinfo{journal}{\emph{Current Paediatrics}} \bibinfo{volume}{14},
  \bibinfo{number}{6} (\bibinfo{year}{2004}), \bibinfo{pages}{482--488}.
\newblock


\bibitem[Karabacak et~al\mbox{.}(2016)]%
        {Karabacak2016}
\bibfield{author}{\bibinfo{person}{Bilge Karabacak},
  \bibinfo{person}{Sevgi~Ozkan Yildirim}, {and} \bibinfo{person}{Nazife
  Baykal}.} \bibinfo{year}{2016}\natexlab{}.
\newblock \showarticletitle{A vulnerability-driven cyber security maturity
  model for measuring national critical infrastructure protection
  preparedness}.
\newblock \bibinfo{journal}{\emph{International Journal of Critical
  Infrastructure Protection}}  \bibinfo{volume}{4} (\bibinfo{date}{Oct}
  \bibinfo{year}{2016}), \bibinfo{pages}{47--59}.
\newblock
\urldef\tempurl%
\url{https://doi.org/10.1016/j.ijcip.2016.10.001}
\showDOI{\tempurl}


\bibitem[Keleba et~al\mbox{.}(2022)]%
        {Keleba2022}
\bibfield{author}{\bibinfo{person}{Habeenzu Keleba}, \bibinfo{person}{Oteng
  Tabona}, {and} \bibinfo{person}{Thabiso Maupong}.}
  \bibinfo{year}{2022}\natexlab{}.
\newblock \bibinfo{title}{Developing a Cyber-resilience state in Botswana's
  Energy Industry}.
\newblock , \bibinfo{numpages}{6}~pages.
\newblock
\urldef\tempurl%
\url{https://doi.org/10.1109/SmartNets55823.2022.9993997}
\showDOI{\tempurl}


\bibitem[Kitchenham(2004)]%
        {Kitchenham04}
\bibfield{author}{\bibinfo{person}{Barbara Kitchenham}.}
  \bibinfo{year}{2004}\natexlab{}.
\newblock \bibinfo{booktitle}{\emph{Procedures for Performing Systematic
  Reviews}}.
\newblock \bibinfo{type}{Keele University. Technical Report TR/SE-0401}.
  \bibinfo{institution}{Keele University}, \bibinfo{address}{Department of
  Computer Science, Keele University, UK}.
\newblock


\bibitem[Kitchenham and Brereton(2013)]%
        {Kitchenham2013}
\bibfield{author}{\bibinfo{person}{Barbara Kitchenham} {and}
  \bibinfo{person}{Pearl Brereton}.} \bibinfo{year}{2013}\natexlab{}.
\newblock \showarticletitle{A systematic review of systematic review process
  research in software engineering}.
\newblock \bibinfo{journal}{\emph{Information and Software Technology}}
  \bibinfo{volume}{55}, \bibinfo{number}{12} (\bibinfo{year}{2013}),
  \bibinfo{pages}{2049--2075}.
\newblock
\showISSN{09505849}
\urldef\tempurl%
\url{https://doi.org/10.1016/j.infsof.2013.07.010}
\showDOI{\tempurl}


\bibitem[Kitchenham and Charters(2007)]%
        {Kitchenham02}
\bibfield{author}{\bibinfo{person}{Barbara Kitchenham} {and}
  \bibinfo{person}{Stuart Charters}.} \bibinfo{year}{2007}\natexlab{}.
\newblock \bibinfo{booktitle}{\emph{Guidelines for performing Systematic
  Literature Reviews in Software Engineering}}.
\newblock \bibinfo{type}{Keele University. ESBE Technical Report EBSE-2007-01}.
  \bibinfo{institution}{Keele University}, \bibinfo{address}{Department of
  Computer Science, Keele University, UK}.
\newblock


\bibitem[Koh et~al\mbox{.}(2011)]%
        {Koh2011}
\bibfield{author}{\bibinfo{person}{Lian~Chee Koh}, \bibinfo{person}{Aidan
  Slingsby}, \bibinfo{person}{Jason Dykes}, {and} \bibinfo{person}{Tin~Seong
  Kam}.} \bibinfo{year}{2011}\natexlab{}.
\newblock \showarticletitle{Developing and applying a user-centered model for
  the design and implementation of information visualization tools}. In
  \bibinfo{booktitle}{\emph{2011 15th International Conference on Information
  Visualisation}}. IEEE, \bibinfo{pages}{90--95}.
\newblock


\bibitem[Konstantinou and Mohanty(2020)]%
        {Konstantinou2020}
\bibfield{author}{\bibinfo{person}{Charalambos Konstantinou} {and}
  \bibinfo{person}{Saraju~P Mohanty}.} \bibinfo{year}{2020}\natexlab{}.
\newblock \showarticletitle{Cybersecurity for the smart grid}.
\newblock \bibinfo{journal}{\emph{Computer}} \bibinfo{volume}{53},
  \bibinfo{number}{5} (\bibinfo{year}{2020}), \bibinfo{pages}{10--12}.
\newblock


\bibitem[Kour et~al\mbox{.}(2020)]%
        {Kour2020}
\bibfield{author}{\bibinfo{person}{Ravdeep Kour}, \bibinfo{person}{Ramin
  Karim}, {and} \bibinfo{person}{Adithya Thaduri}.}
  \bibinfo{year}{2020}\natexlab{}.
\newblock \showarticletitle{Cybersecurity for railways – A maturity model}.
\newblock \bibinfo{journal}{\emph{Proceedings of the Institution of Mechanical
  Engineers, Part F: Journal of Rail and Rapid Transit}} \bibinfo{volume}{234},
  \bibinfo{number}{10} (\bibinfo{year}{2020}), \bibinfo{pages}{1129--1148}.
\newblock
\showISSN{0954-4097, 2041-3017}
\urldef\tempurl%
\url{https://doi.org/10.1177/0954409719881849}
\showDOI{\tempurl}


\bibitem[Kreppein et~al\mbox{.}(2021)]%
        {Kreppein2021}
\bibfield{author}{\bibinfo{person}{Alexander Kreppein},
  \bibinfo{person}{Alexander Kies}, {and} \bibinfo{person}{Robert~H Schmitt}.}
  \bibinfo{year}{2021}\natexlab{}.
\newblock \showarticletitle{Novel Maturity Model for Cybersecurity Evaluation
  in Industry 4.0}. In \bibinfo{booktitle}{\emph{Advances in Cyber Security:
  Third International Conference, ACeS 2021, Penang, Malaysia, August 24--25,
  2021, Revised Selected Papers 3}}. Springer, \bibinfo{pages}{198--210}.
\newblock


\bibitem[Kulugh et~al\mbox{.}(2020)]%
        {Kulugh2022}
\bibfield{author}{\bibinfo{person}{Victor~Emmanuel Kulugh},
  \bibinfo{person}{Uche~M. Mbanaso}, {and} \bibinfo{person}{Gloria
  Chukwudebe}.} \bibinfo{year}{2020}\natexlab{}.
\newblock \showarticletitle{Cybersecurity Resilience Maturity Assessment Model
  for Critical National Information Infrastructure}.
\newblock \bibinfo{journal}{\emph{{SN} Computer Science}} \bibinfo{volume}{3},
  \bibinfo{number}{3} (\bibinfo{year}{2020}), \bibinfo{pages}{217}.
\newblock
\showISSN{2662-995X, 2661-8907}
\urldef\tempurl%
\url{https://doi.org/10.1007/s42979-022-01108-x}
\showDOI{\tempurl}


\bibitem[Le and Hoang(2016)]%
        {Le2016}
\bibfield{author}{\bibinfo{person}{Ngoc~T. Le} {and} \bibinfo{person}{Doan~B.
  Hoang}.} \bibinfo{year}{2016}\natexlab{}.
\newblock \showarticletitle{Can maturity models support cyber security?}. In
  \bibinfo{booktitle}{\emph{2016 {IEEE} 35th International Performance
  Computing and Communications Conference ({IPCCC})}} (Las Vegas, {NV}, {USA},
  2016-12). \bibinfo{publisher}{{IEEE}}, \bibinfo{pages}{1--7}.
\newblock
\showISBNx{978-1-5090-5252-3}
\urldef\tempurl%
\url{https://doi.org/10.1109/PCCC.2016.7820663}
\showDOI{\tempurl}


\bibitem[Le and Hoang(2017)]%
        {Le2017}
\bibfield{author}{\bibinfo{person}{Ngoc~T Le} {and} \bibinfo{person}{Doan~B
  Hoang}.} \bibinfo{year}{2017}\natexlab{}.
\newblock \showarticletitle{Capability maturity model and metrics framework for
  cyber cloud security}.
\newblock \bibinfo{journal}{\emph{Scalable Computing}} (\bibinfo{year}{2017}).
\newblock


\bibitem[Libraries(2017)]%
        {ColoradoState2017}
\bibfield{author}{\bibinfo{person}{Colarado State~University Libraries}.}
  \bibinfo{year}{2017}\natexlab{}.
\newblock \bibinfo{title}{Research Guides: How to Do Library Research:
  Truncation and Proximity Operators}.
\newblock
\newblock
\urldef\tempurl%
\url{https://libguides.colostate.edu/howtodo/truncationproximity#:~:text=The%20truncations%20symbol%20is%20frequently}
\showURL{%
\tempurl}


\bibitem[Lopez-Leyva et~al\mbox{.}(2020)]%
        {Lopez2020}
\bibfield{author}{\bibinfo{person}{Josue~A Lopez-Leyva},
  \bibinfo{person}{Christopher~A Kanter-Ramirez}, {and} \bibinfo{person}{Jose~P
  Morales-Martinez}.} \bibinfo{year}{2020}\natexlab{}.
\newblock \showarticletitle{Customized diagnostic tool for the security
  maturity level of the enterprise information based on ISO/IEC 27001}. In
  \bibinfo{booktitle}{\emph{2020 8th International Conference in Software
  Engineering Research and Innovation (CONISOFT)}}. IEEE,
  \bibinfo{pages}{147--153}.
\newblock


\bibitem[Malatji et~al\mbox{.}(2022)]%
        {Malatji2022}
\bibfield{author}{\bibinfo{person}{Masike Malatji},
  \bibinfo{person}{Annlizé~L. Marnewick}, {and} \bibinfo{person}{Suné
  Von~Solms}.} \bibinfo{year}{2022}\natexlab{}.
\newblock \showarticletitle{Cybersecurity capabilities for critical
  infrastructure resilience}.
\newblock \bibinfo{journal}{\emph{Information \& Computer Security}}
  \bibinfo{volume}{30}, \bibinfo{number}{2} (\bibinfo{date}{Jan}
  \bibinfo{year}{2022}), \bibinfo{pages}{255--279}.
\newblock
\showISSN{2056-4961}
\urldef\tempurl%
\url{https://doi.org/10.1108/ICS-06-2021-0091}
\showDOI{\tempurl}
\newblock
\shownote{Publisher: Emerald Publishing Limited}.


\bibitem[Malatji et~al\mbox{.}(2019)]%
        {Malatji2019}
\bibfield{author}{\bibinfo{person}{Masike Malatji}, \bibinfo{person}{Sune
  Von~Solms}, {and} \bibinfo{person}{Annliz{\'e} Marnewick}.}
  \bibinfo{year}{2019}\natexlab{}.
\newblock \showarticletitle{Socio-technical systems cybersecurity framework}.
\newblock \bibinfo{journal}{\emph{Information \& Computer Security}}
  \bibinfo{volume}{27}, \bibinfo{number}{2} (\bibinfo{year}{2019}),
  \bibinfo{pages}{233--272}.
\newblock


\bibitem[Maleh et~al\mbox{.}(2021)]%
        {Maleh2021}
\bibfield{author}{\bibinfo{person}{Yassine Maleh}, \bibinfo{person}{Abdelkebir
  Sahid}, {and} \bibinfo{person}{Mustapha Belaissaoui}.}
  \bibinfo{year}{2021}\natexlab{}.
\newblock \showarticletitle{A maturity framework for cybersecurity governance
  in organizations}.
\newblock \bibinfo{journal}{\emph{EDPACS}} \bibinfo{volume}{63},
  \bibinfo{number}{6} (\bibinfo{year}{2021}), \bibinfo{pages}{1--22}.
\newblock


\bibitem[Marican et~al\mbox{.}(2023)]%
        {Marican2023}
\bibfield{author}{\bibinfo{person}{Mohamed Noordin~Yusuff Marican},
  \bibinfo{person}{Shukor~Abd Razak}, \bibinfo{person}{Ali Selamat}, {and}
  \bibinfo{person}{Siti~Hajar Othman}.} \bibinfo{year}{2023}\natexlab{}.
\newblock \showarticletitle{Cyber Security Maturity Assessment Framework for
  Technology Startups: A Systematic Literature Review}.
\newblock \bibinfo{journal}{\emph{{IEEE} Access}}  \bibinfo{volume}{11}
  (\bibinfo{year}{2023}), \bibinfo{pages}{5442--5452}.
\newblock
\showISSN{2169-3536}
\urldef\tempurl%
\url{https://doi.org/10.1109/ACCESS.2022.3229766}
\showDOI{\tempurl}


\bibitem[Miron and Muita(2014)]%
        {Walter2014}
\bibfield{author}{\bibinfo{person}{Walter Miron} {and} \bibinfo{person}{Kevin
  Muita}.} \bibinfo{year}{2014}\natexlab{}.
\newblock \showarticletitle{Cybersecurity Capability Maturity Models for
  Providers of Critical Infrastructure}.
\newblock \bibinfo{journal}{\emph{Technology Innovation Management Review}}
  \bibinfo{volume}{4} (\bibinfo{date}{Oct} \bibinfo{year}{2014}),
  \bibinfo{pages}{33--39}.
\newblock
\urldef\tempurl%
\url{https://doi.org/10.22215/timreview/837}
\showDOI{\tempurl}


\bibitem[Mishra(2022)]%
        {Mishra2022}
\bibfield{author}{\bibinfo{person}{Ashish Mishra}.}
  \bibinfo{year}{2022}\natexlab{}.
\newblock \bibinfo{booktitle}{\emph{Modern Cybersecurity Strategies for
  Enterprises: Protect and Secure Your Enterprise Networks, Digital Business
  Assets, and Endpoint Security with Tested and Proven Methods (English
  Edition)}}.
\newblock \bibinfo{publisher}{BPB Publications}.
\newblock


\bibitem[M{\"o}ller(2023)]%
        {Moller2023}
\bibfield{author}{\bibinfo{person}{Dietmar P.~F. M{\"o}ller}.}
  \bibinfo{year}{2023}\natexlab{}.
\newblock \bibinfo{booktitle}{\emph{Cybersecurity Maturity Models and SWOT
  Analysis}}.
\newblock \bibinfo{publisher}{Springer Nature Switzerland},
  \bibinfo{address}{Cham}, \bibinfo{pages}{305--346}.
\newblock
\showISBNx{978-3-031-26845-8}
\urldef\tempurl%
\url{https://doi.org/10.1007/978-3-031-26845-8_7}
\showDOI{\tempurl}


\bibitem[Muronga et~al\mbox{.}(2019)]%
        {Muronga2019}
\bibfield{author}{\bibinfo{person}{Khangwelo Muronga}, \bibinfo{person}{Marlein
  Herselman}, \bibinfo{person}{Adele Botha}, {and} \bibinfo{person}{Ad{\'e}le
  Da~Veiga}.} \bibinfo{year}{2019}\natexlab{}.
\newblock \showarticletitle{An analysis of assessment approaches and maturity
  scales used for evaluation of information security and cybersecurity user
  awareness and training programs: A scoping review}. In
  \bibinfo{booktitle}{\emph{2019 conference on Next Generation Computing
  Applications (NextComp)}}. IEEE, \bibinfo{pages}{1--6}.
\newblock


\bibitem[Page et~al\mbox{.}(2021)]%
        {Page2021}
\bibfield{author}{\bibinfo{person}{Matthew~J. Page}, \bibinfo{person}{Joanne~E.
  McKenzie}, \bibinfo{person}{Patrick~M. Bossuyt}, \bibinfo{person}{Isabelle
  Boutron}, \bibinfo{person}{Tammy~C. Hoffmann}, \bibinfo{person}{Cynthia~D.
  Mulrow}, \bibinfo{person}{Larissa Shamseer}, \bibinfo{person}{Jennifer~M.
  Tetzlaff}, \bibinfo{person}{Elie~A. Akl}, \bibinfo{person}{Sue~E. Brennan},
  \bibinfo{person}{Roger Chou}, \bibinfo{person}{Julie Glanville},
  \bibinfo{person}{Jeremy~M. Grimshaw}, \bibinfo{person}{Asbjørn
  Hróbjartsson}, \bibinfo{person}{Manoj~M. Lalu}, \bibinfo{person}{Tianjing
  Li}, \bibinfo{person}{Elizabeth~W. Loder}, \bibinfo{person}{Evan
  Mayo-Wilson}, \bibinfo{person}{Steve McDonald}, \bibinfo{person}{Luke~A.
  McGuinness}, \bibinfo{person}{Lesley~A. Stewart}, \bibinfo{person}{James
  Thomas}, \bibinfo{person}{Andrea~C. Tricco}, \bibinfo{person}{Vivian~A.
  Welch}, \bibinfo{person}{Penny Whiting}, {and} \bibinfo{person}{David
  Moher}.} \bibinfo{year}{2021}\natexlab{}.
\newblock \showarticletitle{The PRISMA 2020 statement: an updated guideline for
  reporting systematic reviews}.
\newblock \bibinfo{journal}{\emph{Systematic Reviews}}  \bibinfo{volume}{10}
  (\bibinfo{date}{03} \bibinfo{year}{2021}).
\newblock
\urldef\tempurl%
\url{https://doi.org/10.1186/s13643-021-01626-4}
\showDOI{\tempurl}


\bibitem[Prasanna and SaidiReddy(2021)]%
        {Prasanna2021}
\bibfield{author}{\bibinfo{person}{B~Lakshmi Prasanna} {and} \bibinfo{person}{M
  SaidiReddy}.} \bibinfo{year}{2021}\natexlab{}.
\newblock \showarticletitle{(CSM2-RA-R2-TI): Cyber Security Maturity Model for
  Risk Assessment Using Risk Register for Threat Intelligence}. In
  \bibinfo{booktitle}{\emph{Journal of Physics: Conference Series}},
  Vol.~\bibinfo{volume}{2040}. IOP Publishing, \bibinfo{pages}{012005}.
\newblock


\bibitem[Rabii et~al\mbox{.}(2020)]%
        {Rabii2020}
\bibfield{author}{\bibinfo{person}{Anass Rabii}, \bibinfo{person}{Saliha
  Assoul}, \bibinfo{person}{Khadija~Ouazzani Touhami}, {and}
  \bibinfo{person}{Ounsa Roudies}.} \bibinfo{year}{2020}\natexlab{}.
\newblock \showarticletitle{Information and cyber security maturity models: a
  systematic literature review}.
\newblock \bibinfo{journal}{\emph{Information \& Computer Security}}
  \bibinfo{volume}{28}, \bibinfo{number}{4} (\bibinfo{year}{2020}),
  \bibinfo{pages}{627--644}.
\newblock


\bibitem[Rea-Guaman et~al\mbox{.}(2017)]%
        {Rea2017}
\bibfield{author}{\bibinfo{person}{Angel~Marcelo Rea-Guaman},
  \bibinfo{person}{Tom{\'a}s San~Feliu}, \bibinfo{person}{Jose~A.
  Calvo-Manzano}, {and} \bibinfo{person}{Isaac~Daniel Sanchez-Garcia}.}
  \bibinfo{year}{2017}\natexlab{}.
\newblock \showarticletitle{Comparative Study of Cybersecurity Capability
  Maturity Models}. In \bibinfo{booktitle}{\emph{Software Process Improvement
  and Capability Determination}}, \bibfield{editor}{\bibinfo{person}{Antonia
  Mas}, \bibinfo{person}{Antoni Mesquida}, \bibinfo{person}{Rory~V. O'Connor},
  \bibinfo{person}{Terry Rout}, {and} \bibinfo{person}{Alec Dorling}} (Eds.).
  \bibinfo{publisher}{Springer International Publishing},
  \bibinfo{address}{Cham}, \bibinfo{pages}{100--113}.
\newblock
\showISBNx{978-3-319-67383-7}


\bibitem[Reynolds et~al\mbox{.}(2021)]%
        {Reynolds2021}
\bibfield{author}{\bibinfo{person}{Steven~Lamarr Reynolds},
  \bibinfo{person}{Tobias Mertz}, \bibinfo{person}{Steven Arzt}, {and}
  \bibinfo{person}{J{\"o}rn Kohlhammer}.} \bibinfo{year}{2021}\natexlab{}.
\newblock \showarticletitle{User-centered design of visualizations for software
  vulnerability reports}. In \bibinfo{booktitle}{\emph{2021 IEEE Symposium on
  Visualization for Cyber Security (VizSec)}}. IEEE, \bibinfo{pages}{68--78}.
\newblock


\bibitem[Rojas et~al\mbox{.}(2022)]%
        {Rojas2022}
\bibfield{author}{\bibinfo{person}{Aaron Joseph~Serrano Rojas},
  \bibinfo{person}{Erick Fabrizzio~Paniura Valencia}, \bibinfo{person}{Jimmy
  Armas-Aguirre}, {and} \bibinfo{person}{Juan Manuel~Madrid Molina}.}
  \bibinfo{year}{2022}\natexlab{}.
\newblock \showarticletitle{Cybersecurity maturity model for the protection and
  privacy of personal health data}. In \bibinfo{booktitle}{\emph{2022 {IEEE}
  2nd International Conference on Advanced Learning Technologies on Education
  \& Research ({ICALTER})}} (Lima, Peru, 2022-11-16).
  \bibinfo{publisher}{{IEEE}}, \bibinfo{pages}{1--4}.
\newblock
\showISBNx{978-1-66545-696-8}
\urldef\tempurl%
\url{https://doi.org/10.1109/ICALTER57193.2022.9964729}
\showDOI{\tempurl}


\bibitem[Safitra et~al\mbox{.}(2023)]%
        {Safitra2023}
\bibfield{author}{\bibinfo{person}{Muhammad~Fakhrul Safitra},
  \bibinfo{person}{Muharman Lubis}, {and} \bibinfo{person}{Hanif Fakhrurroja}.}
  \bibinfo{year}{2023}\natexlab{}.
\newblock \showarticletitle{Counterattacking cyber threats: A framework for the
  future of cybersecurity}.
\newblock \bibinfo{journal}{\emph{Sustainability}} \bibinfo{volume}{15},
  \bibinfo{number}{18} (\bibinfo{year}{2023}), \bibinfo{pages}{13369}.
\newblock


\bibitem[Santos-Neto and Costa(2019)]%
        {Santos2019}
\bibfield{author}{\bibinfo{person}{João Batista Sarmento~dos Santos-Neto}
  {and} \bibinfo{person}{Ana Paula Cabral~Seixas Costa}.}
  \bibinfo{year}{2019}\natexlab{}.
\newblock \showarticletitle{Enterprise maturity models: a systematic literature
  review}.
\newblock \bibinfo{journal}{\emph{Enterprise Information Systems}}
  \bibinfo{volume}{13}, \bibinfo{number}{5} (\bibinfo{year}{2019}),
  \bibinfo{pages}{719--769}.
\newblock
\showISSN{1751-7575, 1751-7583}
\urldef\tempurl%
\url{https://doi.org/10.1080/17517575.2019.1575986}
\showDOI{\tempurl}


\bibitem[Sava{\c{s}} and Karata{\c{s}}(2022)]%
        {Savacs2022}
\bibfield{author}{\bibinfo{person}{Serkan Sava{\c{s}}} {and}
  \bibinfo{person}{S{\"u}leyman Karata{\c{s}}}.}
  \bibinfo{year}{2022}\natexlab{}.
\newblock \showarticletitle{Cyber governance studies in ensuring cybersecurity:
  an overview of cybersecurity governance}.
\newblock \bibinfo{journal}{\emph{International Cybersecurity Law Review}}
  \bibinfo{volume}{3}, \bibinfo{number}{1} (\bibinfo{year}{2022}),
  \bibinfo{pages}{7--34}.
\newblock


\bibitem[Schlette et~al\mbox{.}(2021)]%
        {Schlette2021}
\bibfield{author}{\bibinfo{person}{Daniel Schlette}, \bibinfo{person}{Manfred
  Vielberth}, {and} \bibinfo{person}{G{\"u}nther Pernul}.}
  \bibinfo{year}{2021}\natexlab{}.
\newblock \showarticletitle{CTI-SOC2M2--The quest for mature,
  intelligence-driven security operations and incident response capabilities}.
\newblock \bibinfo{journal}{\emph{Computers \& Security}}
  \bibinfo{volume}{111} (\bibinfo{year}{2021}), \bibinfo{pages}{102482}.
\newblock


\bibitem[Selamat et~al\mbox{.}(2022)]%
        {Selamat2022}
\bibfield{author}{\bibinfo{person}{Ali Selamat}, \bibinfo{person}{Mohamed
  Noordin~Yusuff Marican}, \bibinfo{person}{Siti~Hajar Othman}, {and}
  \bibinfo{person}{Shukor Abd~Razak}.} \bibinfo{year}{2022}\natexlab{}.
\newblock \showarticletitle{An End-To-End Cyber Security Maturity Model For
  Technology Startups}. In \bibinfo{booktitle}{\emph{2022 IEEE International
  Conference on Computing (ICOCO)}}. IEEE, \bibinfo{pages}{185--190}.
\newblock


\bibitem[Shaked et~al\mbox{.}(2020)]%
        {Shaked2020}
\bibfield{author}{\bibinfo{person}{Avi Shaked}, \bibinfo{person}{Lior
  Tabansky}, {and} \bibinfo{person}{Yoram Reich}.}
  \bibinfo{year}{2020}\natexlab{}.
\newblock \showarticletitle{Incorporating systems thinking into a cyber
  resilience maturity model}.
\newblock \bibinfo{journal}{\emph{IEEE Engineering Management Review}}
  \bibinfo{volume}{49}, \bibinfo{number}{2} (\bibinfo{year}{2020}),
  \bibinfo{pages}{110--115}.
\newblock


\bibitem[Sharkov(2020)]%
        {Sharkov2020}
\bibfield{author}{\bibinfo{person}{George Sharkov}.}
  \bibinfo{year}{2020}\natexlab{}.
\newblock \showarticletitle{Assessing the maturity of national cybersecurity
  and resilience}.
\newblock \bibinfo{journal}{\emph{Connections: The Quarterly Journal}}
  \bibinfo{volume}{19}, \bibinfo{number}{4} (\bibinfo{year}{2020}),
  \bibinfo{pages}{5--24}.
\newblock


\bibitem[Shillair et~al\mbox{.}(2022)]%
        {Shillair2022}
\bibfield{author}{\bibinfo{person}{Ruth Shillair}, \bibinfo{person}{Patricia
  Esteve-Gonz{\'a}lez}, \bibinfo{person}{William~H Dutton},
  \bibinfo{person}{Sadie Creese}, \bibinfo{person}{Eva Nagyfejeo}, {and}
  \bibinfo{person}{Basie von Solms}.} \bibinfo{year}{2022}\natexlab{}.
\newblock \showarticletitle{Cybersecurity education, awareness raising, and
  training initiatives: National level evidence-based results, challenges, and
  promise}.
\newblock \bibinfo{journal}{\emph{Computers \& Security}}
  \bibinfo{volume}{119} (\bibinfo{year}{2022}), \bibinfo{pages}{102756}.
\newblock


\bibitem[Skarga-Bandurova et~al\mbox{.}(2021)]%
        {Skarga2021}
\bibfield{author}{\bibinfo{person}{Inna Skarga-Bandurova},
  \bibinfo{person}{Igor Kotsiuba}, {and} \bibinfo{person}{Erkuden~Rios
  Velasco}.} \bibinfo{year}{2021}\natexlab{}.
\newblock \showarticletitle{Cyber Hygiene Maturity Assessment Framework for
  Smart Grid Scenarios}.
\newblock \bibinfo{journal}{\emph{Frontiers in Computer Science}}
  \bibinfo{volume}{3} (\bibinfo{year}{2021}), \bibinfo{pages}{614337}.
\newblock


\bibitem[Sony and Naik(2020)]%
        {Sony2020}
\bibfield{author}{\bibinfo{person}{Michael Sony} {and} \bibinfo{person}{Subhash
  Naik}.} \bibinfo{year}{2020}\natexlab{}.
\newblock \showarticletitle{Industry 4.0 integration with socio-technical
  systems theory: A systematic review and proposed theoretical model}.
\newblock \bibinfo{journal}{\emph{Technology in Society}}  \bibinfo{volume}{61}
  (\bibinfo{year}{2020}), \bibinfo{pages}{101248}.
\newblock
\showISSN{0160-791X}
\urldef\tempurl%
\url{https://doi.org/10.1016/j.techsoc.2020.101248}
\showDOI{\tempurl}


\bibitem[Spruit and R{\"o}ling(2014)]%
        {Spruit2014}
\bibfield{author}{\bibinfo{person}{Marco~Ren{\'e} Spruit} {and}
  \bibinfo{person}{M.W.M. R{\"o}ling}.} \bibinfo{year}{2014}\natexlab{}.
\newblock \showarticletitle{ISFAM: The Information Security Focus Area Maturity
  Model}. In \bibinfo{booktitle}{\emph{European Conference on Information
  Systems}}.
\newblock
\urldef\tempurl%
\url{https://api.semanticscholar.org/CorpusID:14568926}
\showURL{%
\tempurl}


\bibitem[Srivastava et~al\mbox{.}(2024)]%
        {Srivastava2024}
\bibfield{author}{\bibinfo{person}{Durgesh Srivastava},
  \bibinfo{person}{Rajeshwar Singh}, \bibinfo{person}{Chinmay Chakraborty},
  \bibinfo{person}{Sunil~Kr Maakar}, \bibinfo{person}{Aaisha Makkar}, {and}
  \bibinfo{person}{Deepak Sinwar}.} \bibinfo{year}{2024}\natexlab{}.
\newblock \showarticletitle{A framework for detection of cyber attacks by the
  classification of intrusion detection datasets}.
\newblock \bibinfo{journal}{\emph{Microprocessors and Microsystems}}
  \bibinfo{volume}{105} (\bibinfo{year}{2024}), \bibinfo{pages}{104964}.
\newblock


\bibitem[van Dinter et~al\mbox{.}(2021)]%
        {Van2021}
\bibfield{author}{\bibinfo{person}{Raymon van Dinter}, \bibinfo{person}{Bedir
  Tekinerdogan}, {and} \bibinfo{person}{Cagatay Catal}.}
  \bibinfo{year}{2021}\natexlab{}.
\newblock \showarticletitle{Automation of systematic literature reviews: A
  systematic literature review}.
\newblock \bibinfo{journal}{\emph{Information and Software Technology}}
  \bibinfo{volume}{136} (\bibinfo{year}{2021}), \bibinfo{pages}{106589}.
\newblock
\showISSN{09505849}
\urldef\tempurl%
\url{https://doi.org/10.1016/j.infsof.2021.106589}
\showDOI{\tempurl}


\bibitem[Ventures(2023)]%
        {Ventures2023}
\bibfield{author}{\bibinfo{person}{Cybersecurity Ventures}.}
  \bibinfo{year}{2023}\natexlab{}.
\newblock \bibinfo{title}{2023 Official Cybercrime Report}.
\newblock
\newblock
\urldef\tempurl%
\url{https://www.esentire.com/resources/library/2023-official-cybercrime-report}
\showURL{%
\tempurl}


\bibitem[von Solms(2015)]%
        {Von2015}
\bibfield{author}{\bibinfo{person}{SH~Basie von Solms}.}
  \bibinfo{year}{2015}\natexlab{}.
\newblock \showarticletitle{A maturity model for part of the African Union
  Convention on Cyber Security}. In \bibinfo{booktitle}{\emph{2015 Science and
  Information Conference (SAI)}}. IEEE, \bibinfo{pages}{1316--1320}.
\newblock


\bibitem[W(2018)]%
        {Anne2018}
\bibfield{author}{\bibinfo{person}{Anne W}.} \bibinfo{year}{2018}\natexlab{}.
\newblock \bibinfo{title}{Maturity models in cyber security: what's happening
  to the IAMM?}
\newblock
\newblock
\urldef\tempurl%
\url{https://www.ncsc.gov.uk/blog-post/maturity-models-cyber-security-whats-happening-iamm}
\showURL{%
\tempurl}


\bibitem[Wallace et~al\mbox{.}(2021)]%
        {Wallace2021}
\bibfield{author}{\bibinfo{person}{Steven Wallace}, \bibinfo{person}{Karen
  Green}, \bibinfo{person}{Catherine Johnson}, \bibinfo{person}{Joseph Cooper},
  {and} \bibinfo{person}{Collin Gilstrap}.} \bibinfo{year}{2021}\natexlab{}.
\newblock \showarticletitle{An extended TOE framework for cybersecurity
  adoption decisions}.
\newblock \bibinfo{journal}{\emph{Communications of the Association for
  Information Systems}} \bibinfo{volume}{47}, \bibinfo{number}{2020}
  (\bibinfo{year}{2021}), \bibinfo{pages}{51}.
\newblock


\bibitem[Wang et~al\mbox{.}(2018)]%
        {Wang2018}
\bibfield{author}{\bibinfo{person}{Lihui Wang}, \bibinfo{person}{Xi~Vincent
  Wang}, \bibinfo{person}{Lihui Wang}, {and} \bibinfo{person}{Xi~Vincent
  Wang}.} \bibinfo{year}{2018}\natexlab{}.
\newblock \showarticletitle{Challenges in Cybersecurity}.
\newblock \bibinfo{journal}{\emph{Cloud-Based Cyber-Physical Systems in
  Manufacturing}} (\bibinfo{year}{2018}), \bibinfo{pages}{63--79}.
\newblock


\bibitem[White(2011)]%
        {White2011}
\bibfield{author}{\bibinfo{person}{Gregory~B. White}.}
  \bibinfo{year}{2011}\natexlab{}.
\newblock \showarticletitle{The community cyber security maturity model}. In
  \bibinfo{booktitle}{\emph{2011 {IEEE} International Conference on
  Technologies for Homeland Security ({HST})}} (Waltham, {MA}, {USA}).
  \bibinfo{publisher}{{IEEE}}, \bibinfo{pages}{173--178}.
\newblock
\showISBNx{978-1-4577-1376-7 978-1-4577-1375-0 978-1-4577-1374-3}
\urldef\tempurl%
\url{https://doi.org/10.1109/THS.2011.6107866}
\showDOI{\tempurl}


\bibitem[Whitworth(2009)]%
        {Whitworth2009}
\bibfield{author}{\bibinfo{person}{Brian Whitworth}.}
  \bibinfo{year}{2009}\natexlab{}.
\newblock \showarticletitle{A brief introduction to sociotechnical systems}.
\newblock In \bibinfo{booktitle}{\emph{Encyclopedia of Information Science and
  Technology, Second Edition}}. \bibinfo{publisher}{IGI Global},
  \bibinfo{pages}{394--400}.
\newblock


\bibitem[Yigit~Ozkan and Spruit(2023)]%
        {Yigit2023}
\bibfield{author}{\bibinfo{person}{Bilge Yigit~Ozkan} {and}
  \bibinfo{person}{Marco Spruit}.} \bibinfo{year}{2023}\natexlab{}.
\newblock \showarticletitle{Adaptable Security Maturity Assessment and
  Standardization for Digital SMEs}.
\newblock \bibinfo{journal}{\emph{Journal of Computer Information Systems}}
  \bibinfo{volume}{63}, \bibinfo{number}{4} (\bibinfo{year}{2023}),
  \bibinfo{pages}{965--987}.
\newblock


\bibitem[Yigit~Ozkan et~al\mbox{.}(2020)]%
        {Yigit2020}
\bibfield{author}{\bibinfo{person}{Bilge Yigit~Ozkan}, \bibinfo{person}{Marco
  Spruit}, \bibinfo{person}{Roland Wondolleck}, {and} \bibinfo{person}{Veronica
  Burriel~Coll}.} \bibinfo{year}{2020}\natexlab{}.
\newblock \showarticletitle{Modelling adaptive information security for SMEs in
  a cluster}.
\newblock \bibinfo{journal}{\emph{Journal of Intellectual Capital}}
  \bibinfo{volume}{21}, \bibinfo{number}{2} (\bibinfo{year}{2020}),
  \bibinfo{pages}{235--256}.
\newblock


\bibitem[Yigit~Ozkan et~al\mbox{.}(2021)]%
        {Yigit2021}
\bibfield{author}{\bibinfo{person}{Bilge Yigit~Ozkan}, \bibinfo{person}{Sonny
  Van~Lingen}, {and} \bibinfo{person}{Marco Spruit}.}
  \bibinfo{year}{2021}\natexlab{}.
\newblock \showarticletitle{The {Cybersecurity} {Focus} {Area} {Maturity}
  ({CYSFAM}) {Model}}.
\newblock \bibinfo{journal}{\emph{Journal of Cybersecurity and Privacy}}
  \bibinfo{volume}{1}, \bibinfo{number}{1} (\bibinfo{date}{Feb.}
  \bibinfo{year}{2021}), \bibinfo{pages}{119--139}.
\newblock
\showISSN{2624-800X}
\urldef\tempurl%
\url{https://doi.org/10.3390/jcp1010007}
\showDOI{\tempurl}


\bibitem[Yulianto et~al\mbox{.}(2016)]%
        {Yulianto2016}
\bibfield{author}{\bibinfo{person}{Semi Yulianto}, \bibinfo{person}{Charles
  Lim}, {and} \bibinfo{person}{Benfano Soewito}.}
  \bibinfo{year}{2016}\natexlab{}.
\newblock \showarticletitle{Information security maturity model: A best
  practice driven approach to PCI DSS compliance}. In
  \bibinfo{booktitle}{\emph{2016 IEEE Region 10 Symposium (TENSYMP)}}. IEEE,
  \bibinfo{pages}{65--70}.
\newblock


\end{thebibliography}

\end{document}